\documentclass[12pt,english,prd,nofootinbib,preprint,floatfix,tightenlines]{revtex4}
\usepackage[T1]{fontenc}
\usepackage[latin1]{inputenc}
\usepackage{array}
\usepackage{verbatim}
\usepackage{graphicx}
\usepackage{amssymb}
\usepackage{here}

\makeatletter

\providecommand{\tabularnewline}{\\}

\usepackage{babel}
\makeatother
\begin{document}
\setcounter{footnote}{0}

\preprint{ANL-PR-07-17, MSUHEP-080125, NSF-KITP-08-26, UCRHEP-T447, arXiv:0802.0007 {[}hep-ph]}

\title{Implications of CTEQ global analysis for collider observables}

\author{Pavel M. Nadolsky,$^{1}$ Hung-Liang Lai,$^{2,3}$ Qing-Hong Cao,$^{4}$
Joey Huston,$^{1}$ Jon Pumplin,$^{1}$ Daniel Stump,$^{1}$ Wu-Ki
Tung,$^{1,3}$ and C.-P. Yuan$^{1}$}

\affiliation{$^{1}$Department of Physics and Astronomy, Michigan State University,
East Lansing, MI 48824-1116, U.S.A.\\
 $^{2}$Taipei Municipal University of Education, Taipei, Taiwan\\
$^{3}$Department of Physics, University of Washington, Seattle, WA
98105, U.S.A.\\
 $^{4}$Department of Physics and Astronomy, University of California
at Riverside, Riverside, CA 92521, U.S.A. }

\date{March 17, 2008}

\begin{abstract}
 The latest CTEQ6.6 parton distributions, obtained by global analysis
of hard scattering data in the framework of general-mass perturbative
QCD, are employed to study theoretical predictions and their uncertainties
for significant processes at the Fermilab Tevatron and CERN Large
Hadron Collider. The previously observed increase in predicted cross
sections for the standard-candle $W$ and $Z$ boson production processes
in the general-mass scheme (compared to those in the zero-mass scheme)
is further investigated and quantified. A novel method to constrain
PDF uncertainties in LHC observables, by effectively exploiting PDF-induced
correlations with benchmark standard model cross sections, is presented.
Using this method, we show that the $t\bar{t}$ cross section can
potentially serve as a standard candle observable for the LHC processes
dominated by initial-state gluon scattering. Among other benefits,
precise measurements of $t\bar{t}$ cross sections would reduce PDF
uncertainties in predictions for single-top quark and Higgs boson
production in the standard model and minimal supersymmetric standard
model.
\end{abstract}



\maketitle
\tableofcontents{}

\newpage

\section{Introduction\label{sec:Introduction}}

Parton distribution functions (PDFs) are essential inputs required
to make theoretical predictions for the CERN Large Hadron Collider
(LHC) and other hadron scattering facilities. They are extracted from
a comprehensive global analysis of hard-scattering data from a variety
of fixed-target and collider experiments in the framework of perturbative
QCD. Experimental groups envision an ambitious program to tightly
constrain PDF degrees of freedom using the upcoming LHC data. Such
constraints will not be feasible in the early runs of the LHC, when
experimental systematic errors will typically be large, and the collider
luminosity itself will not be known to better than $10-20\%$. Thus
the experiments plan to perform real-time monitoring of the collider
luminosity through the measurement of benchmark standard model cross
sections, notably those for production of massive electroweak bosons
\cite{Dittmar:1997md,Khoze:2000db,Giele:2001ms,Frisch1994}. These
cross sections are large and can be measured fairly precisely soon
after the LHC turn-on. To realize this goal, as well as to carry out
the general physics program of the LHC, it is important, on one hand,
to systematically explore the dependence of $W$, $Z$, and other
{}``standard candle'' cross sections on the PDFs and other aspects
of QCD theory; and, on the other hand, to establish and exploit \emph{correlations}
with these observables arising from the dependence on the universal
parton distributions. These tasks must be carried out using the most
up-to-date PDFs, with quantitative estimates of their uncertainties.

In a series of recent papers \cite{Tung:2006tb,Lai:2007dq,Pumplin:2007wg},
we have extended the conventional CTEQ global PDF analysis \cite{Pumplin:2002vw,Stump:2003yu}
to incorporate a comprehensive treatment of heavy-quark effects and
to include the most recent experimental data. \textbf{}The PDFs constructed
in those studies consist of (i) the base set CTEQ6.5M, together with
40 eigenvector sets along 20 orthonormal directions in the parton
parameter space \cite{Tung:2006tb}; (ii) several PDF sets CTEQ6.5Sn
(n=-2,...4), designed to probe the strangeness degrees of freedom
under the assumption of symmetric or asymmetric strange sea \cite{Lai:2007dq};
and (iii) several sets CTEQ6.5XCn (n=0...6) for a study of the charm
sector of the parton parameter space, in particular, the allowed range
of independent nonperturbative ({}``intrinsic'') charm partons in
several possible models \cite{Pumplin:2007wg}. Some prominent physical
consequences of these new PDFs compared to previous PDFs, particularly
for $W$ and $Z$ production cross sections at hadron colliders, were
pointed out in these papers.

The current paper pursues a more detailed exploration of the physical
implications of our new generation of PDFs, particularly at the LHC.
In the process of this detailed study, we found it desirable to expand
and improve the CTEQ6.5 analysis on several fronts. All new results
presented in this paper are based on these improved PDFs, which we
designate as CTEQ6.6.%
\footnote{The CTEQ6.6 PDFs represent improvements over their counterparts in
CTEQ6.5, while preserving the same physics inputs. Therefore we recommend
that CTEQ6.6 be the preferred PDFs to use in future phenomenology
studies. %
} They will be described in Sec.~\ref{sec:PDFplots}.

We make a systematic effort to address the quantitative challenges
described in the first paragraph of this introduction. We study the
impact of the new PDFs on the predicted cross sections for important
physical processes at the LHC, with their associated uncertainty ranges.
The PDF uncertainties in future measurements may behave at odds with
initial intuitions because of rich connections between PDFs of different
flavors and in different kinematical ranges arising from (a) physical
symmetries, such as scale invariance and parton sum rules, and (b)
experimental constraints. In some cases, notably in $Z$ boson production
at the LHC, the largest PDF uncertainty arises from less-constrained
subleading scattering channels rather than from the well-known dominant
subprocesses. In order to access such rich interconnections efficiently
and completely, we introduce in Sec.~\ref{sec:Correlation-analysis}
a quantitative measure of correlations existing between the PDFs and
physical observables. The correlation analysis extends the Hessian
method \cite{Pumplin:2001ct,Pumplin:2002vw} to investigate pairwise
relations between collider observables. This analysis is employed
in Sec\@.~\ref{sec:ImplicationsForColliders} as a tool to enhance
the power of certain crucial phenomenological predictions, notably
for $W$ and $Z$ boson production cross sections, and to examine
the role of $t\bar{t}$ production as an additional standard candle
process. At the end of Sec.~\ref{sec:ImplicationsForColliders},
we apply the gained knowledge to identify the main sources of PDF
uncertainty in single-top production and in several processes for
Higgs boson production in the standard model (SM) and minimal supersymmetric
standard model (MSSM).



\section{Correlations due to PDFs\label{sec:Correlation-analysis}}

In many applications, it is instructive to establish whether a collider
observable shares common degrees of freedom with precisely measured
SM cross sections through the nonperturbative PDF parametrizations,
a feature that can be exploited to predict the observable more reliably.
In Section~\ref{sec:ImplicationsForColliders}, we will explore such
PDF-induced correlations between interesting collider cross
sections. But first we will define the relevant theoretical framework.

Let $X$ be a variable that depends on the PDFs. It can be any one
of the physical quantities of interest, or even a PDF itself at some
given $(x,\mu)$. We consider $X$ as a function of the parameters
$\{ a_{i}\}$ that define the PDFs at the initial scale $\mu_{0}$.
Thus we have $X(\vec{a})$, where $\vec{a}$ forms a vector in an
$N$-dimensional PDF parameter space, with $N$ being the number of
free parameters in the global analysis that determines these PDFs.
In the Hessian formalism for the uncertainty analysis developed in
\cite{Pumplin:2001ct} and used in all of our recent work, this parton
parameter space is spanned by a set of orthonormal eigenvectors obtained
by a self-consistent iterative procedure \cite{Tung:2006tb,Pumplin:2002vw}.

If $\vec{a}_{0}$ represents the best fit obtained with a given set
of theoretical and experimental inputs, the variation of $X(\vec{a})$
for parton parameters $\vec{a}$ in the neighborhood of $\vec{a}_{0}$
is given, within the Hessian approximation, by a linear formula\begin{equation}
\Delta X(\vec{a})=X(\vec{a})-X(\vec{a}_{0})=\vec{\nabla}X|_{\vec{a}_{0}}\cdot\Delta\vec{a},\end{equation}
 where $\vec{\nabla}X$ is the gradient of $X(\vec{a}),$ and $\Delta\vec{a}=\vec{a}-\vec{a}_{0}$.
As explained in detail in Refs.~\cite{Pumplin:2001ct,Pumplin:2002vw,Tung:2006tb},
the uncertainty range of the PDFs in our global analysis is characterized
by a tolerance factor $T$, equal to the radius of a hypersphere spanned
by maximal allowed displacements $\Delta\vec{a}$ in the orthonormal
PDF parameter representation. $T$ is determined by the criterion
that all PDFs within this tolerance hypersphere should be consistent
with the input experimental data sets within roughly 90\% c.l. The
detailed discussions and the specific iterative procedure used to
construct the eigenvectors can be found in Refs.~\cite{Pumplin:2001ct,Pumplin:2002vw,Tung:2006tb}.

In practice, the results of our uncertainty analysis are characterized
by $2N$ sets of published eigenvector PDF sets along with the central
fit. We have $2$ PDF sets for each of the $N$ eigenvectors, along
the $(\pm)$ directions respectively, at the distance $|\Delta\vec{a}|=T$.
The $i$-th component of the gradient vector $\vec{\nabla}X$ may
be approximated by\begin{equation}
\frac{\partial X}{\partial a_{i}}\equiv\partial_{i}X=\frac{1}{2}(X_{i}^{(+)}-X_{i}^{(-)}),\label{dXdzi}\end{equation}
 where $X_{i}^{(+)}$ and $X_{i}^{(-)}$ are the values of $X$ computed
from the two sets of PDFs along the ($\pm$) direction of the $i$-th
eigenvector. The uncertainty of the quantity $X$ due to its dependence
on the PDFs is then defined as\begin{equation}
\Delta X=\left\vert \vec{\nabla}X\right\vert =\frac{1}{2}\sqrt{\sum_{i=1}^{N}\left(X_{i}^{(+)}-X_{i}^{(-)}\right)^{2}},\label{masterDX}\end{equation}
 where for simplicity we assume that the positive and negative errors
on $X$ are the same.%
\footnote{A more detailed equation for $\Delta X$ accounts for differences
between the positive and negative errors \cite{Nadolsky:2001yg,Stump:2003yu}.
It is used for $t\bar{t}$ cross sections in Table~\ref{tab:ttbarABC}
and Fig.~\ref{fig:ttbarLHC}. %
}

\begin{figure}
\includegraphics[width=1\columnwidth]{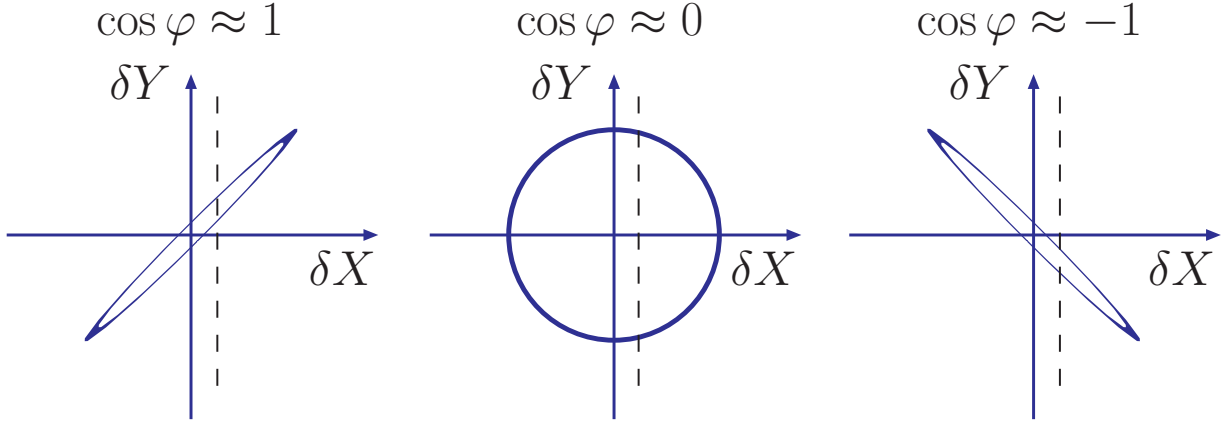}

\caption{Dependence on the correlation ellipse formed in the $\delta X-\delta Y$
plane on the value of $\cos\varphi.$ }

\label{fig:CorrelationEllipsePhi} 
\end{figure}

We may extend the uncertainty analysis to define a \emph{correlation}
between the uncertainties of two variables, say $X(\vec{a})$ and
$Y(\vec{a}).$ We consider the projection of the tolerance hypersphere
onto a circle of radius 1 in the plane of the gradients $\vec{\nabla}X$
and $\vec{\nabla}Y$ in the parton parameter space \cite{Pumplin:2001ct,Nadolsky:2001yg}.
The circle maps onto an ellipse in the $XY$ plane. This {}``tolerance
ellipse'' is described by Lissajous-style parametric equations,\begin{eqnarray}
X & = & X_{0}+\Delta X\cos\theta,\label{ellipse1}\\
Y & = & Y_{0}+\Delta Y\cos(\theta+\varphi),\label{ellipse2}\end{eqnarray}
 where the parameter $\theta$ varies between 0 and $2\pi$, $X_{0}\equiv X(\vec{a}_{0}),$
and $Y_{0}\equiv Y(\vec{a}_{0})$. $\Delta X$ and $\Delta Y$ are
the maximal variations $\delta X\equiv X-X_{0}$ and $\delta Y\equiv Y-Y_{0}$
evaluated according to Eq.~(\ref{masterDX}), and $\varphi$ is the
angle between $\vec{\nabla}X$ and $\vec{\nabla}Y$ in the $\{ a_{i}\}$
space, with\begin{equation}
\cos\varphi=\frac{\vec{\nabla}X\cdot\vec{\nabla}Y}{\Delta X\Delta Y}=\frac{1}{4\Delta X\,\Delta Y}\sum_{i=1}^{N}\left(X_{i}^{(+)}-X_{i}^{(-)}\right)\left(Y_{i}^{(+)}-Y_{i}^{(-)}\right).\label{cosphi}\end{equation}

The quantity $\cos\varphi$ characterizes whether the PDF degrees
of freedom of $X$ and $Y$ are correlated ($\cos\varphi\approx1$),
anti-correlated ($\cos\varphi\approx-1$), or uncorrelated ($\cos\varphi\approx0$).
If units for $X$ and $Y$ are rescaled so that $\Delta X=\Delta Y$
(e.g., $\Delta X=\Delta Y=1$), the semimajor axis of the tolerance
ellipse is directed at an angle $\pi/4$ (or $3\pi/4)$ with respect
to the $\Delta X$ axis for $\cos\varphi>0$ (or $\cos\varphi<0$).
In these units, the ellipse reduces to a line for $\cos\varphi=\pm1$
and becomes a circle for $\cos\varphi=0$, as illustrated by Fig.~\ref{fig:CorrelationEllipsePhi}.
These properties can be found by diagonalizing the equation for the
correlation ellipse,%
\begin{equation}
\left(\frac{\delta X}{\Delta X}\right)^{2}+\left(\frac{\delta Y}{\Delta Y}\right)^{2}-2\left(\frac{\delta X}{\Delta X}\right)\left(\frac{\delta Y}{\Delta Y}\right)\cos\varphi=\sin^{2}\varphi.\label{ellipse3}\end{equation}

A magnitude of $|\cos\varphi|$ close to unity suggests that a precise
measurement of $X$ (constraining $\delta X$ to be along the dashed
line in Fig.~\ref{fig:CorrelationEllipsePhi}) is likely to constrain
tangibly the uncertainty $\delta Y$ in $Y$, as the value of $Y$
shall lie within the needle-shaped error ellipse. Conversely, $\cos\varphi\approx0$
implies that the measurement of $X$ is not likely to constrain $\delta Y$
strongly.%
\footnote{The allowed range of $\delta Y/\Delta Y$ for a given $\delta\equiv\delta X/\Delta X$
is $r_{Y}^{(-)}\leq\delta Y/\Delta Y\leq r_{Y}^{(+)},$ where $r_{Y}^{(\pm)}\equiv\delta\cos\varphi\pm\sqrt{1-\delta^{2}}\sin\varphi.$%
} 

The parameters of the correlation ellipse are sufficient to deduce,
under conventional approximations, a Gaussian probability distribution
$P(X,Y|\mbox{CTEQ6.6})$ for finding certain values of $X$ and $Y$
based on the pre-LHC data sets included in the CTEQ6.6 analysis. If
the LHC measures $X$ and $Y$ nearly independently of the PDF model,
a new confidence region for \emph{$X$} and \emph{$Y$} satisfying
both the CTEQ6.6 and LHC constraints can be determined by combining
the prior probability $P(X,Y|\mbox{CTEQ6.6})$ with the new probability
distribution $P(X,Y|\mbox{LHC})$ provided by the LHC measurement\emph{.}
For this purpose, it suffices to construct a probability distribution\begin{equation}
P(X,Y|\mbox{CTEQ6.6+LHC})=P(X,Y|\mbox{CTEQ6.6})P(X,Y|\mbox{LHC}),\label{PCTEQLHC}\end{equation}
which establishes the combined CTEQ6.6+LHC confidence region without
repeating the global fit. 

The values of $\Delta X,$ $\Delta Y,$ and $\cos\varphi$ are also
sufficient to estimate the PDF uncertainty of any function $f(X,Y)$
of $X$ and $Y$ by relating the gradient of $f(X,Y)$ to $\partial_{X}f\equiv\partial f/\partial X$
and $\partial_{Y}f\equiv\partial f/\partial Y$ via the chain rule:\begin{equation}
\Delta f=\left|\vec{\nabla}f\right|=\sqrt{\left(\Delta X\ \partial_{X}f\ \right)^{2}+2\Delta X\ \Delta Y\ \cos\varphi\ \partial_{X}f\ \partial_{Y}f+\left(\Delta Y\ \partial_{Y}f\right)^{2}}.\label{df}\end{equation}
Of particular interest is the case of a rational function $f(X,Y)=X^{m}/Y^{n},$
pertinent to computations of various cross section ratios, cross section
asymmetries, and statistical significance for finding signal events
over background processes \cite{Nadolsky:2001yg}. For rational functions
Eq.~(\ref{df}) takes the form\begin{equation}
\frac{\Delta f}{f_{0}}=\sqrt{\left(m\frac{\Delta X}{X_{0}}\ \right)^{2}-2mn\frac{\Delta X}{X_{0}}\ \frac{\Delta Y}{Y_{0}}\ \cos\varphi\ +\left(n\frac{\Delta Y\ }{Y_{0}}\right)^{2}}.\label{dfrat}\end{equation}
For example, consider a simple ratio, $f=X/Y$. Then $\Delta f/f_{0}$
is suppressed ($\Delta f/f_{0}\approx\left|\Delta X/X_{0}-\Delta Y/Y_{0}\right|$)
if $X$ and $Y$ are strongly correlated, and it is enhanced ($\Delta f/f_{0}\approx\Delta X/X_{0}+\Delta Y/Y_{0}$)
if $X$ and $Y$ are strongly anticorrelated.

As would be true for any estimate provided by the Hessian method,
the correlation angle is inherently approximate. Eq.~(\ref{cosphi})
is derived under a number of simplifying assumptions, notably in the
quadratic approximation for the $\chi^{2}$ function within the tolerance
hypersphere, and by using a symmetric finite-difference formula (\ref{dXdzi})
for $\{\partial_{i}X\}$ that may fail if $X$ is not monotonic. With
these limitations in mind, we find the correlation angle to be a convenient
measure of interdependence between quantities of diverse nature, such
as physical cross sections and parton distributions themselves. For
collider applications, the correlations between measured cross sections
for crucial SM and beyond SM processes will be of primary interest,
as we shall illustrate in Sec.~\ref{sec:ImplicationsForColliders}.
As a first example however, we shall present some representative results
on correlations between the PDFs in the next section.

\section{Overview of CTEQ6.6 PDFs\label{sec:PDFplots}}

\subsection{CTEQ6.6 versus CTEQ6.1 \label{sub:CTEQ6.6-versus-CTEQ6.1}}

The CTEQ6.5 PDFs \cite{Tung:2006tb,Lai:2007dq,Pumplin:2007wg} and
their improved version CTEQ6.6 presented here are based on a new implementation
of heavy-quark mass effects in perturbative QCD cross sections, realized 
in the ACOT
general-mass (GM) variable number scheme \cite{Collins:1998rz,Aivazis:1993pi}
and supplemented by a unified treatment of both kinematical 
and dynamical effects according to the modern 
 SACOT \cite{Collins:1998rz,Kramer:2000hn} and  ACOT-$\chi$
\cite{Tung:2001mv} concepts.
This improvement leads to significant changes in some key predictions
compared to the ordinary zero-mass (ZM) scheme.  The
quality of the global analysis is further enhanced
by to the inclusion of newer data sets, replacement of structure functions
$F_{2}(x,Q)$ and $F_{3}(x,Q)$ by the directly measured cross sections
in H1 and ZEUS deep-inelastic scattering (DIS) data sets, relaxation
of ad hoc constraints on the parametrization of strange quark PDFs,
and improvements in the global fitting procedure. The CTEQ6.6 PDF
set includes four additional PDF eigenvectors to accommodate the free
strangeness parametrization, as described below. It is also in a
better agreement with HERA charm production cross sections than
CTEQ6.5. The public distributions,
available at the LHAPDF depository \cite{LHAPDF}, include
a central PDF set, denoted as CTEQ6.6M, and 44  eigenvector sets 
that span the range of uncertainties in the parton parameter space 
due to input experimental errors.

We illustrate the impact of the improved treatment of heavy-quark
scattering by comparing predictions made using the CTEQ6.6 PDFs (realized
in the GM scheme) and zero-mass CTEQ6.1 PDFs \cite{Stump:2003yu}
(realized in the ZM scheme). All of the cross sections in our global
analysis are calculated at next-to-leading order (NLO) in perturbative
QCD. Figures showing comparison of CTEQ6.6 and CTEQ6.1 for various
PDF flavors are collected at \cite{CTEQ66WebsitePDFs}. Since these
figures are rather similar to their counterparts comparing the CTEQ6.5
and CTEQ6.1 PDFs \cite{Tung:2006tb}, we do not reproduce them in
this paper, except for the strangeness and charm PDFs.

Incorporation of the full
heavy-quark mass effects in the general-mass formalism 
leads to the suppression of heavy-flavor contributions to
the DIS structure functions $F_{\lambda}(x,Q)$ compared to
the zero-mass formalism. For neutral-current  $F_{\lambda}(x,Q)$, 
which dominate the global analysis, 
the suppressions occur in (a) the flavor-excitation partonic processes
with incoming $c$ and $b$ quarks,
through the rescaling of the light-cone momentum fraction variable;
and (b) the light-flavor scattering processes involving 
explicit flavor creation (such as the gluon-photon fusion), 
through the mass dependence in the partonic cross section (Wilson
coefficient) \cite{Tung:2001mv,Tung:2006tb}. Since the theoretical
calculations in the global fit must agree with the extensive DIS
data at low and moderate $Q$, the reduction in $c$, $b$, and $g$ 
contributions in the GM NLO fit must be compensated by larger
magnitudes of light quark and antiquark contributions. One therefore
expects to see an increase in the light-quark PDFs
extracted from CTEQ6.6 compared to those from CTEQ6.1 analyses in
the appropriate $(x,Q)$ region. 

The most consequential differences between CTEQ6.6 and CTEQ6.1 PDFs for 
$u$ and $d$ quarks occur at $x\lesssim10^{-3}$,
cf. Ref.~\cite{CTEQ66WebsitePDFs}. They can substantially affect
predictions for quark-induced processes at the LHC, making the
predicted cross sections larger by several percent 
(6-7\% in $W$, $Z$ production), as will be discussed
in more detail in Sec.~\ref{sec:ImplicationsForColliders}.
%
\begin{figure}
\begin{centering}\includegraphics[clip,width=0.49\columnwidth]{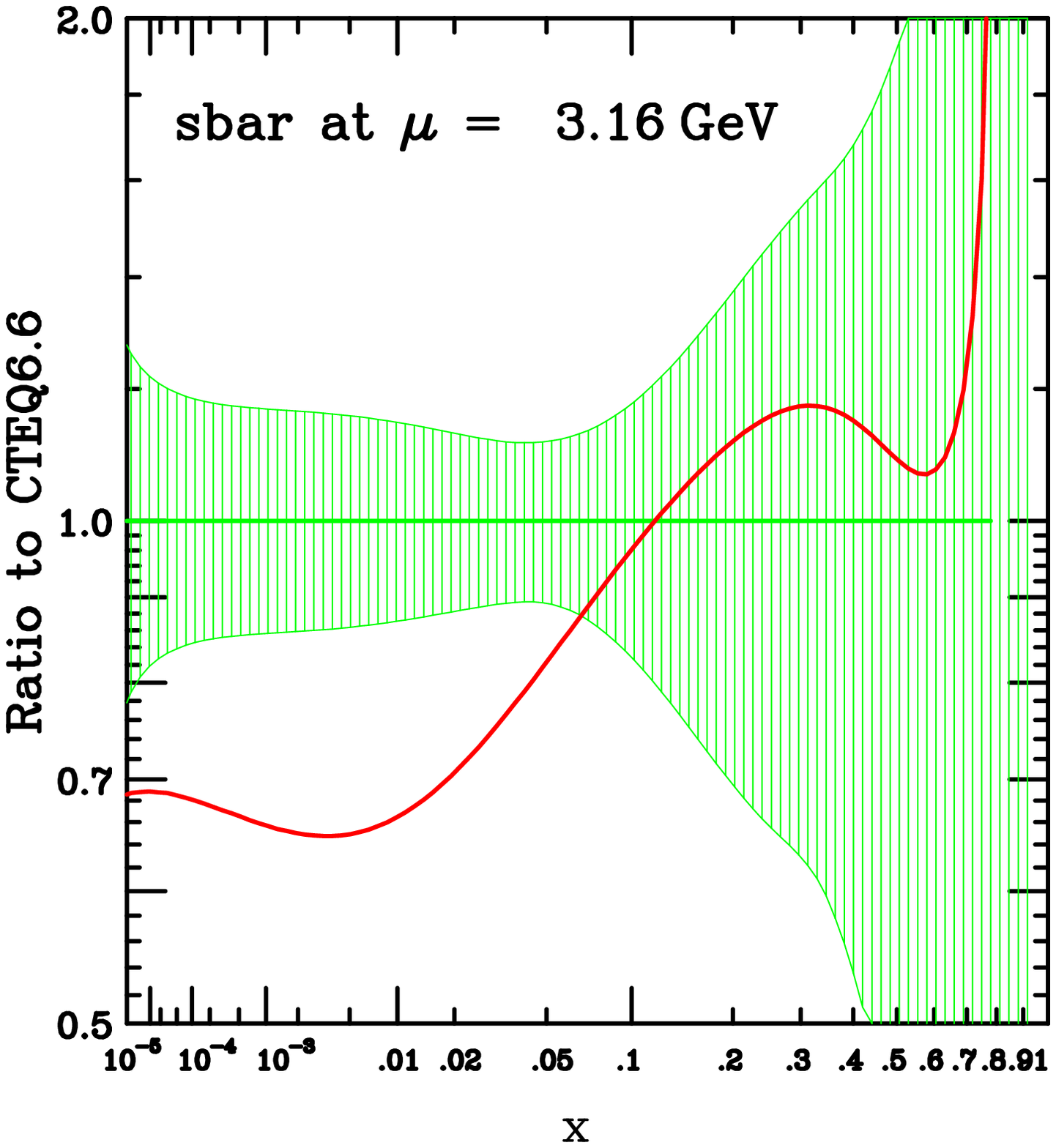}
\includegraphics[clip,width=0.49\columnwidth]{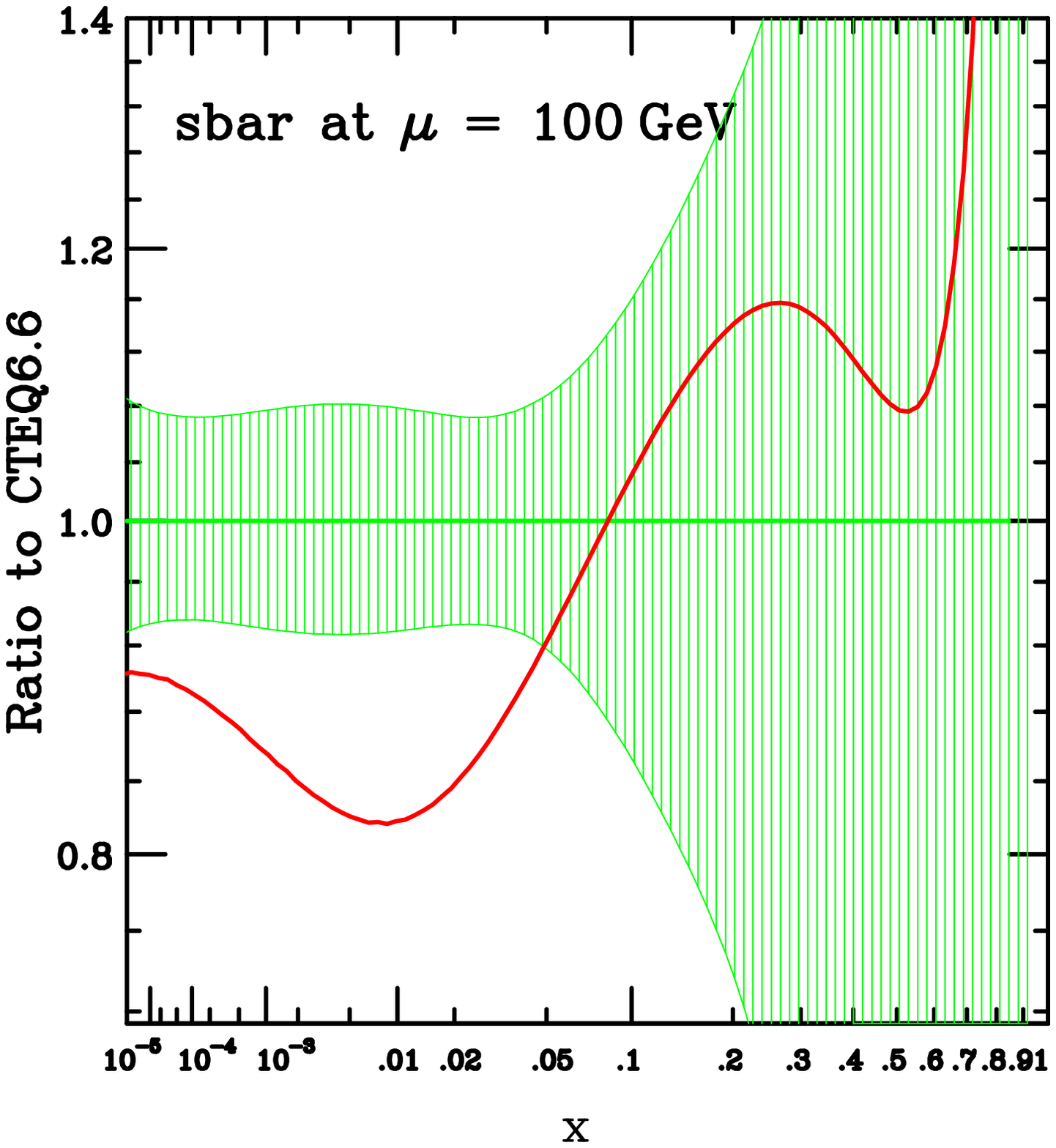} \par\end{centering}

\caption{CTEQ6.6 PDF uncertainty bands (green shaded area) and CTEQ6.1M PDF
(red solid line) for $s=\bar{s}$ at factorization scales $\mu=3.16$
and $100$ GeV.\label{fig:CTEQ66strange}}
\end{figure}

As a new feature, the CTEQ6.6
analysis allows the shape of strange quark distributions to be
independent from the non-strange sea distributions. We no longer
impose the familiar ansatz
$s(x,\mu_{0})\propto\bar{u}(x,\mu_{0})+\bar{d}(x,\mu_{0})$, because
the included dimuon DIS data ($\nu A\to\mu^{+}\mu^{-}X$)
\cite{Tzanov:2005kr} probes the strange quark distributions via the
underlying process $sW\to c$, making the above ansatz unnecessary.
However, as shown in Ref.\,\cite{Lai:2007dq}, the existing
experimental constraints on the strange PDFs remain relatively weak
and have power to determine at most two new degrees of freedom
associated with the strangeness in the limited range $x>10^{-2}$. Thus in the
CTEQ6.6 analysis we add two new free parameters characterizing the
strange PDFs.\footnote{Assignment of more than
two free strangeness parameters does not tangibly improve
the quality of the fit and creates undesirable flat directions in
the Hessian eigenvector space \cite{Lai:2007dq}.}
We continue to assume $s(x)=\bar{s}(x)$ in these fits,
since the current data do not place statistically significant
constraints on the difference between $s(x)$ and $\bar{s}(x)$
\cite{Lai:2007dq}.

At $x\lesssim 10^{-2}$, the available data probes mostly a
combination $(4/9)\left[u(x)+\bar u(x)\right] + (1/9)\left[d(x)+\bar
d(x)+s(x)+\bar s(x)\right]$ accessible in neutral-current DIS, but
not the detailed flavor composition of the quark sea. The shape of
$s(x,\mu_{0})$ at very small $x$ can vary over a large range, 
accompanied by corresponding adjustments 
in the other sea quark flavors. In other
words, the strangeness to non-strangeness ratio at small $x$,
$R_s=\lim_{x\rightarrow
0}\left[s(x,\mu_{0})/\left(\bar{u}(x,\mu_{0})+\bar{d}(x,\mu_{0})\right)\right]$,
is entirely unconstrained by the data. But, on general physics grounds,
one would expect this ratio to be of order 1 (or, arguably, a bit
smaller). Thus, in the current CTEQ6.6 analysis, we adopt a
parametrization for the strange PDF of the form
$s(x,\mu_{0})=A_{0}\, x^{A_{1}}\,(1-x)^{A_{2}}P(x)$, where  
$A_{1}$ is set equal to the analogous parameter of $\bar{u}$
and $\bar{d}$ based on Regge considerations.
A smooth function $P(x)$ (of a fixed form for all 45 CTEQ6.6 PDF sets)
is chosen to ensure that the ratio $R_s$
stays within a reasonable range. There is considerable freedom in
the choice of $P(x)$; and that is a part of the theoretical
uncertainty associated with any parametrization of the initial
PDFs.\footnote{The theoretical uncertainty associated with the
small-$x$ quark flavor composition is generally larger than the Hessian
uncertainty bands due to propagation of experimental errors 
for a given $P(x)$.} 
Fig.~\ref{fig:CTEQ66strange} shows
the CTEQ6.6 uncertainty bands of $\bar{s}(x,\mu)$ at two values of
$\mu$, along with the corresponding CTEQ6.1M PDFs. The values of
$R_s$ for the 44 eigenvector sets in this PDF series span 0.63 -
1.15.

Since the PDFs are de facto used for ultra-high energy and
astrophysics applications at $x<10^{-5}$, the CTEQ6.6 PDFs are
tabulated down to $x=10^{-8}$ to provide numerically stable PDF
values obtained by QCD evolution from parametrized initial parton
distributions in this extreme $x$ region. At the initial scale
$\mu_{0},$ the PDFs are extrapolated into the region $x<10^{-5}$,
not covered by the fitted data, by a Regge-like functional form
proportional to $x^{a}$, where the negative parameter $a$ is found
from the fit at $x>10^{-5}$. At larger $\mu$ scales, the PDFs are
predicted from the initial condition at $\mu=\mu_{0}$ based on the
NLO DGLAP evolution. No extra dynamical effects that may be significant
at small $x$ are included in this parametrization.

\begin{figure}
\begin{centering}\includegraphics[clip,width=0.5\textwidth]{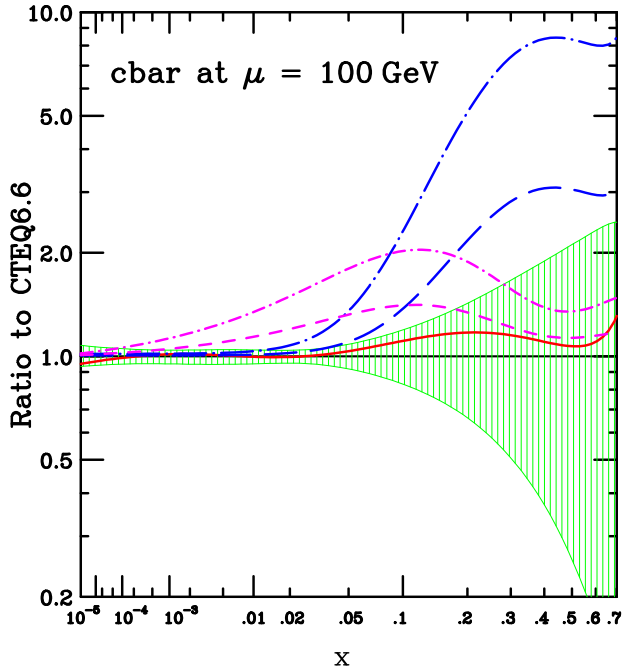} \par\end{centering}

\caption{CTEQ6.6 PDF uncertainty band for $c=\bar{c}$ with radiatively generated
charm only (green shaded area) and PDFs with intrinsic charm: BHPS
form with moderate (long-dashed) and strong (long-dash-dotted) IC;
sea-like form with moderate (short-dashed) and strong (short-dash-dotted)
IC. The red solid line is for the CTEQ6.1M PDF. A factorization scale
$\mu=100$ GeV is assumed.\label{fig:CTEQ66IC}}
\end{figure}

\subsection{Fits with nonperturbative charm}

In the general-purpose CTEQ6.6 PDFs, we assume there is no nonperturbative intrinsic
charm (IC), so that $c(x,\mu_{0})=\bar{c}(x,\mu_{0})=0$ at the initial
evolution scale $\mu_{0}=m_{c}=1.3\,\mathrm{GeV}$. To facilitate
studies of the effect of possible IC, we have also created fits in
which various amounts of IC (with $c=\bar{c}$) are assumed. These
fits amount to updated versions of the light-cone motivated (BHPS \cite{Brodsky:1980pb})
and the sea-like models for the shape of the input charm PDF discussed
in \cite{Pumplin:2007wg}. For each IC model, we provide two PDF parameterizations
with moderate and strong IC contributions, corresponding to charm
quarks carrying 1\% and 3.5\% of the parent nucleon's momentum at
$\mu=m_{c}$, respectively. Figure~\ref{fig:CTEQ66IC} shows that
the assumption of IC can substantially increase the amount of $c=\bar{c}$
at factorization scales as large as $100\,\mathrm{GeV}$.

\begin{figure}
\begin{centering}\includegraphics[width=0.5\columnwidth,height=8cm,keepaspectratio]{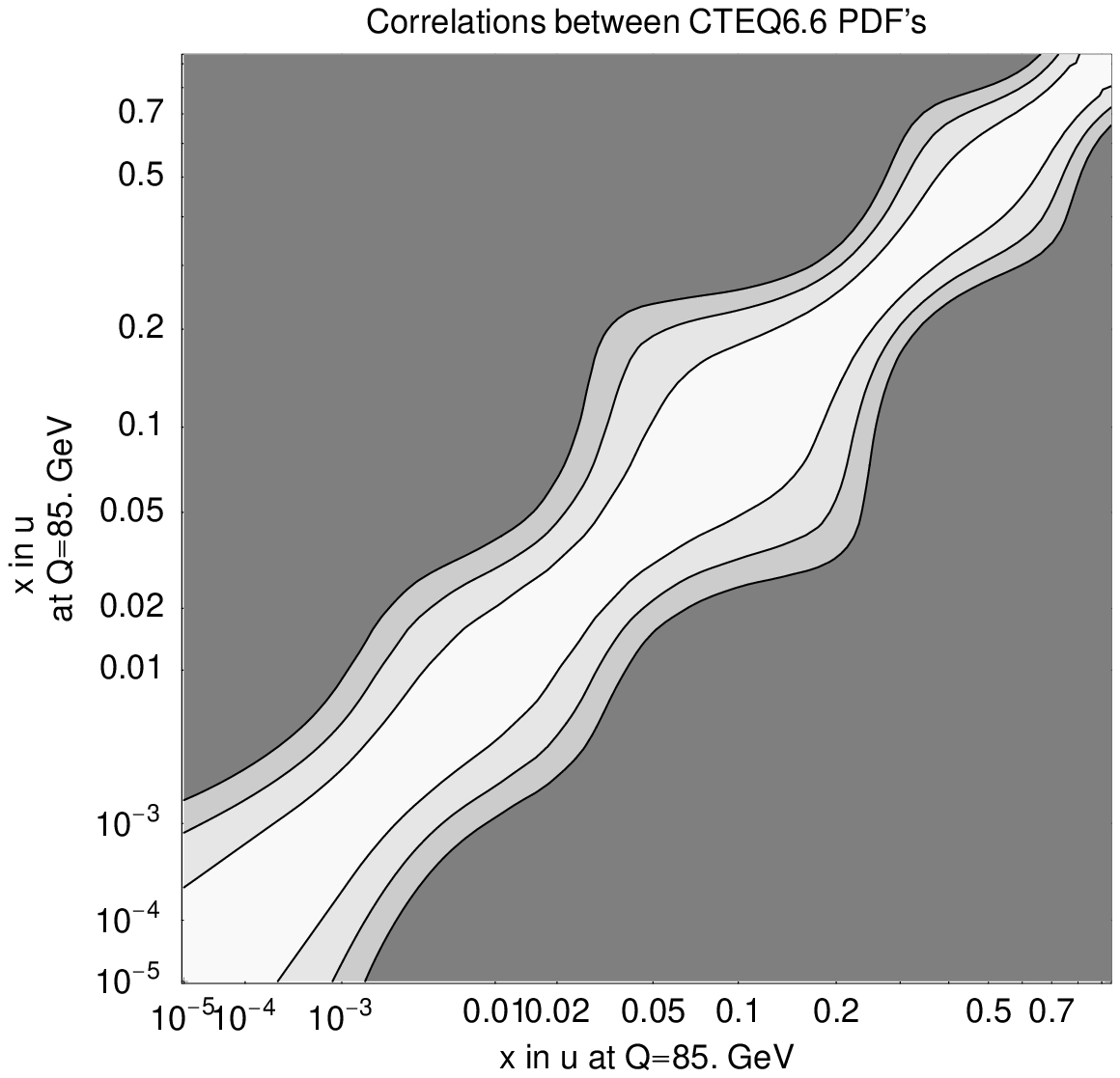}\includegraphics[width=0.5\columnwidth,height=8cm,keepaspectratio]{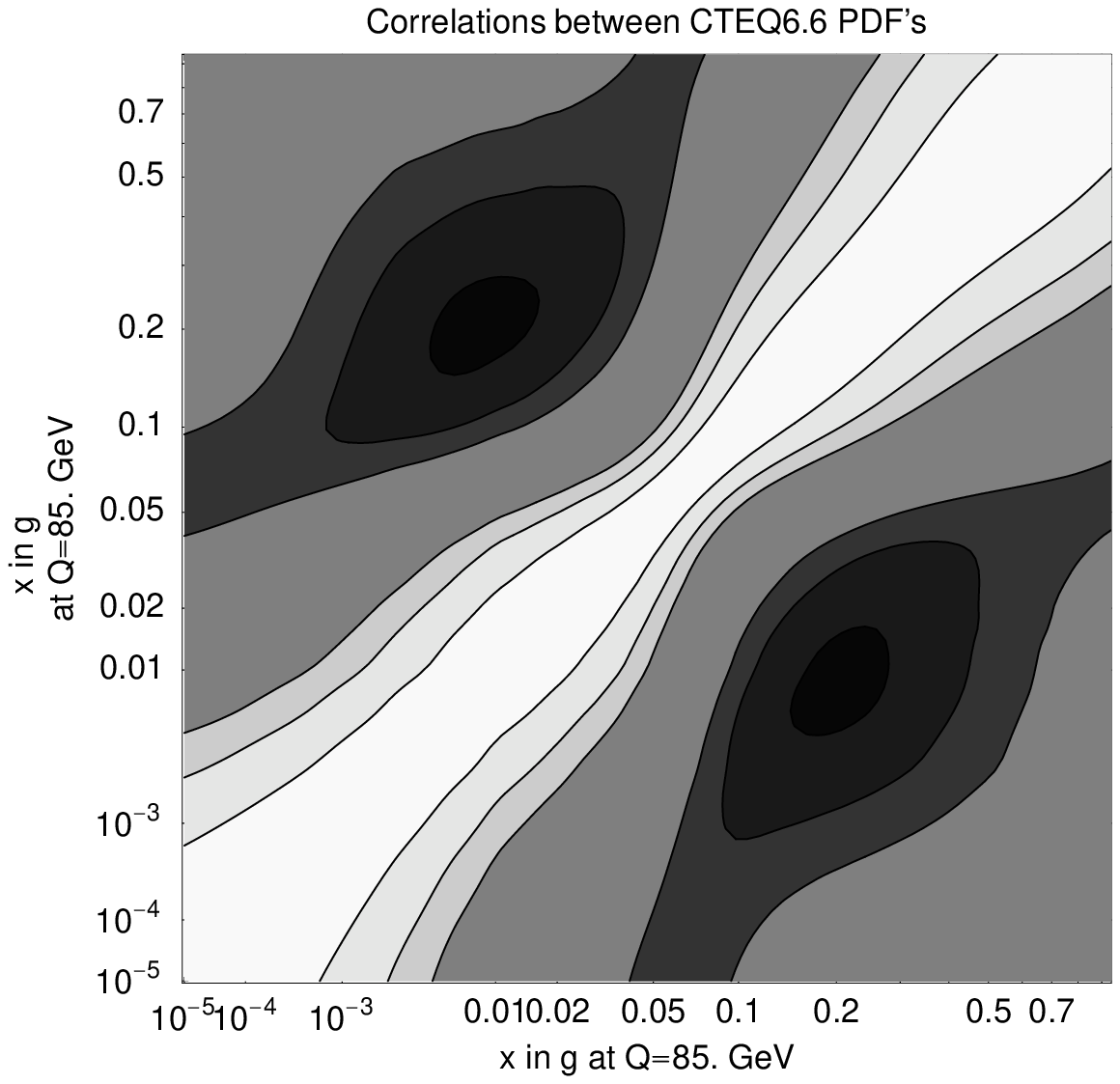}\\
(a)\hspace{3in} (b)\par\end{centering}

\begin{centering}\includegraphics[width=0.5\columnwidth,height=8cm,keepaspectratio]{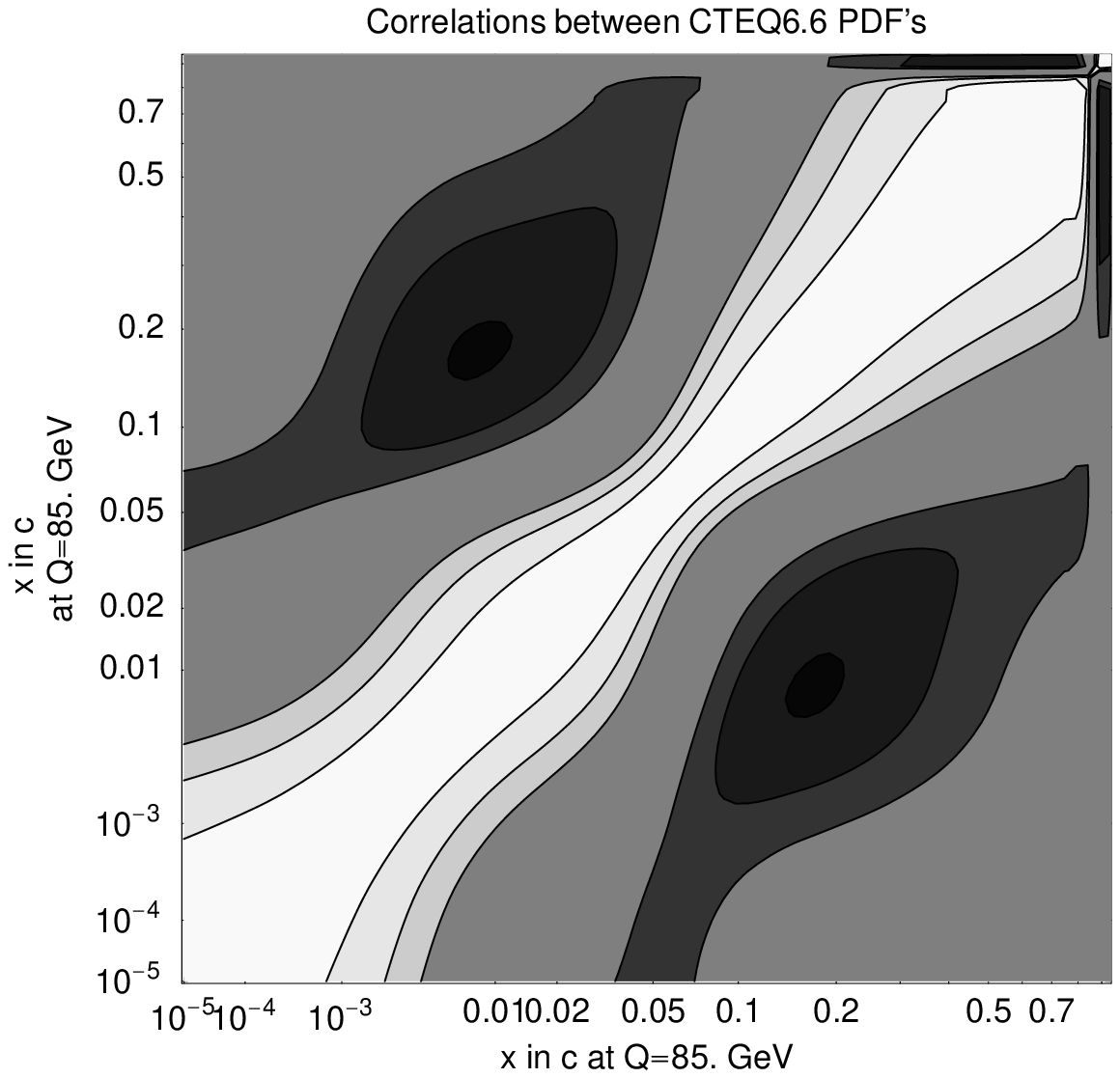}\includegraphics[width=0.5\columnwidth,height=8cm,keepaspectratio]{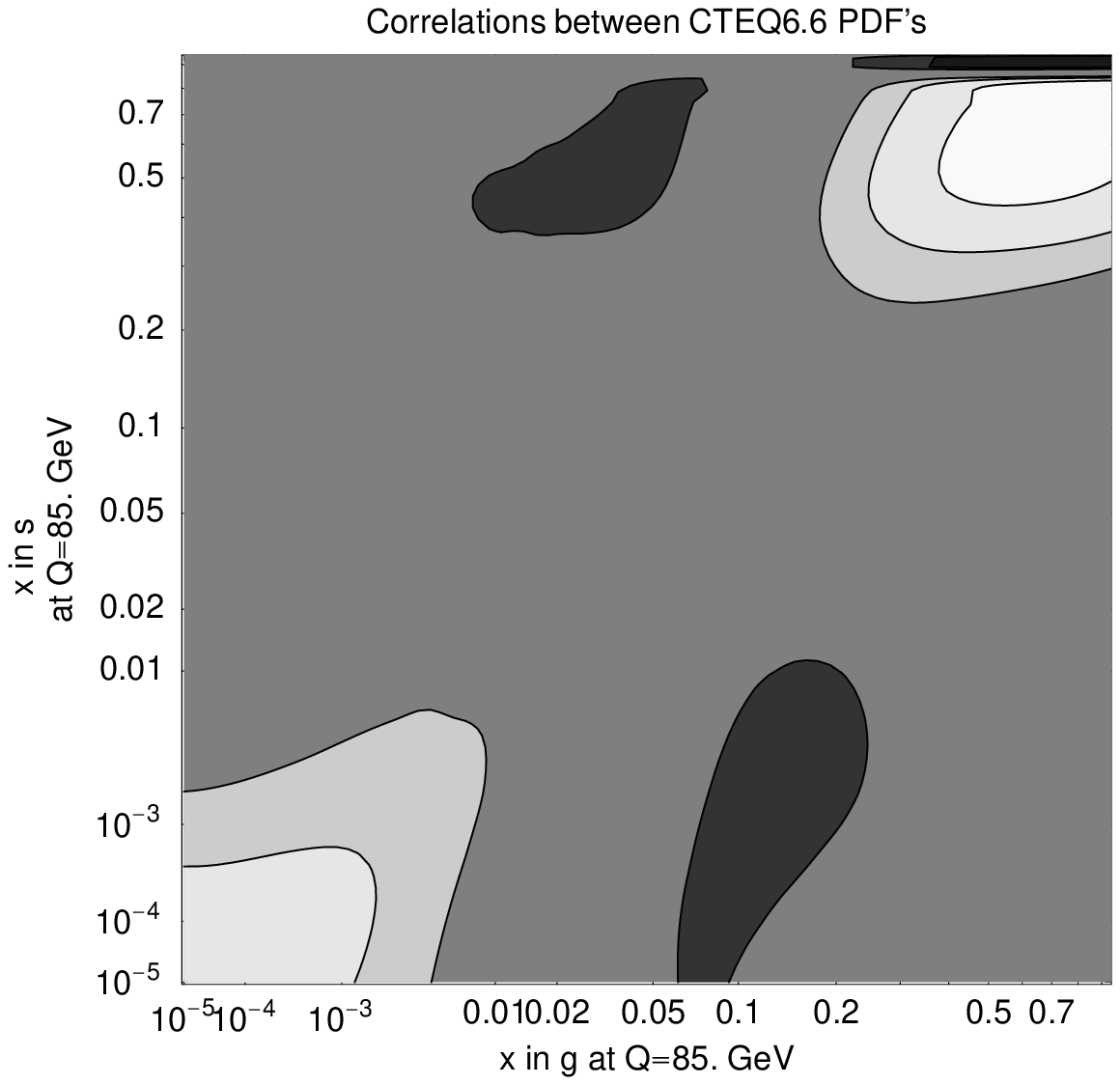}\par\end{centering}

\begin{centering}(c)\hspace{3in} (d)\par\end{centering}

\includegraphics[width=0.3\textwidth]{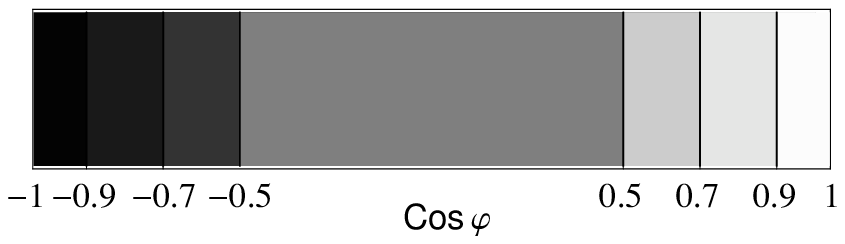}

\caption{Contour plots of correlations $\cos\varphi$ between two PDFs $f_{1}(x_{1},\mu)$
and $f_{2}(x_{2},\mu)$ at $\mu=85\mbox{ GeV}$: (a) $u-u$; (b) $g-g$;
(c) $c-c$; (d) $s-g$. Both axes are scaled as $x^{0.2}$. \label{fig:ContourPlots}}
\end{figure}

\subsection{Correlations\label{sub:PDFCorrelations}}

As mentioned before, correlations can be computed for any variables,
including the PDFs themselves. The $x$ and $\mu$ dependence of the PDF-PDF
correlations provides broad insights about theoretical and experimental
constraints affecting physical cross sections. To explore this dependence,
we present contour plots of the correlation cosine $\cos\varphi$
for pairs of PDFs $f_{a_{1}}(x_{1},\mu_{1})$ and $f_{a_{2}}(x_{2},\mu_{2})$
at scales $\mu_{1,2}=2$ and 85 GeV, plotted as a function of momentum
fractions $x_{1}$ and $x_{2}$. A few such plots relevant for the
ensuing discussion are shown in Fig.~\ref{fig:ContourPlots}. A complete
set of the contour plots in color and grayscale versions is available
at \cite{CTEQ66WebsitePDFCorrs}.

Light (dark) shades of the gray color in Fig.~\ref{fig:ContourPlots}
correspond to magnitudes of $\cos\varphi$ close to 1 (-1), as indicated
by the legend. Several interesting patterns of the PDF correlations
can be generally observed. First, consider self-correlations, in which
the correlation cosine is formed between two values of the same PDF
($a_{1}=a_{2}$, $\mu_{1}=\mu_{2}$) evaluated at momentum fractions
$x_{1}$ and $x_{2}$. The examples include the $u-u,$ $g-g,$ and
$c-c$ correlations at $\mu_{1}=\mu_{2}=85$~GeV shown in Figs.~\ref{fig:ContourPlots}(a)-(c).

Each self-correlation plot includes a trivial correlation, $\cos\varphi\approx1,$
occurring when $x_{1}$ and $x_{2}$ are about the same. This correlation
occurs in light-colored areas along the $x_{1}=x_{2}$ diagonals in
Figs.~\ref{fig:ContourPlots}(a)-(c), of the shape that depends
on the flavor of the PDF and the associated $\mu$ scale.

In the case of an up quark {[}Fig.~\ref{fig:ContourPlots}(a)], the
trivial correlation is the only pronounced pattern visible in the
contour plot. The gluon PDF {[}Fig.~\ref{fig:ContourPlots}(b)] and
related $c,$ $b$ PDFs {[}Fig.~\ref{fig:ContourPlots}(c)] show
an additional strong anti-correlation in the vicinity of $(x_{1},x_{2})\approx(0.2,0.01)$
arising as a consequence of the momentum sum rule. Important implications
of this anticorrelation will be discussed in Section~\ref{sec:ImplicationsForColliders}.

The gluon-strangeness correlation in Fig.~\ref{fig:ContourPlots}(d)
illustrates some typical patterns encountered in the case of sea partons.
The gluons show a correlated behavior with strange quarks (and generally,
sea quarks) at $x<10^{-4}-10^{-3}$ as a reflection of the singlet
evolution (a light area in the lower left corner). The gluon PDF is
anticorrelated with the strangeness PDF at $x_{1}\sim0.1,$ $x_{2}<0.01$
because of the momentum sum rule (a dark area at the bottom). This
anticorrelation significantly affects predictions for $W$ and $Z$
cross sections at the LHC. More complicated (and not so well-understood)
patterns occur at $x_{1,2}>0.3,$ where the sea-parton behavior is
less constrained by the data. Other types of correlation patterns,
associated with the sum rules, perturbative evolution, and constraints
from the experimental data, can be observed in the full set of contour
plots \cite{CTEQ66WebsitePDFCorrs}.

\section{Implications for collider experiments\label{sec:ImplicationsForColliders}}

Soon after its turn-on, the LHC will provide vast samples of data
in well-understood scattering processes at the electroweak scale,
notably production of massive weak bosons $W^{\pm}$ and $Z^{0}$.
These data will facilitate useful experimental calibrations as well
as measurements of the LHC luminosity and PDFs with tentative experimental
accuracy of about 1\% \cite{Dittmar:1997md,Khoze:2000db,Giele:2001ms}.
To utilize such measurements of {}``standard candle'' cross sections
most productively, one must understand \emph{how} they constrain PDF
degrees of freedom. It is necessary to explore how the predicted cross
sections change due to improvements in the theoretical model, such
as the transition from the zero-mass to general-mass factorization
scheme, as well as due to the remaining freedom in the PDF parameters
allowed by the data in the global fit. 

In this section, we examine the quantitative connections between the
PDFs and physical observables by applying the correlation analysis
introduced in Section~\ref{sec:Correlation-analysis}. Since only
large correlations (or anticorrelations) will be useful for practical
purposes, we concentrate on pairs of cross sections characterized
by large magnitudes of the correlation cosine, chosen in this paper
to be $\left|\cos\varphi\right|>0.7.$ Overall dependence on the heavy-quark
scheme is explored in Section~\ref{sub:TotalXSecsTevLHC}. A focused
study of total inclusive cross sections is presented for $W,$ $Z,$
top-quark, and Higgs boson production in Sections~\ref{sub:W-and-Z},
\ref{sub:Top-quark-production}, and \ref{sub:ttbLuminosityHiggs}.
The total cross sections are computed at the next-to-leading
order (NLO) in the QCD coupling strength $\alpha_{s}$, using $W$
boson mass $M_{W}=80.403$ GeV, $Z$ boson mass $M_{Z}=91.1876$ GeV,
and top quark mass $m_{t}=171$ GeV. 

In all calculations, both the
renormalization and factorization scales are set to be equal to the
mass of the final-state heavy particle, unless specified otherwise.
The results presented below are representative of generic patterns observed 
in the PDF dependence of the studied cross sections. They depend weakly 
on theoretical assumptions about the quark flavor composition 
at small $x$ discussed in Section~\ref{sec:PDFplots}. 

%

The LHC collaborations intend to measure ratios of LHC cross sections
to {}``standard-candle'' (especially $W$ and $Z$) cross sections.
If two cross sections $X$ and $Y$ share common systematics, both
experimental and theoretical, the systematic uncertainties partially
cancel in the ratio $r=X/Y$. This cancellation is especially important
in the first year or two of the LHC running, when the collider luminosity
will only be known to the order of $10-20\%$. Eventually the LHC
ratios can provide important precision tests of the Standard Model,
such as a precise measurement of the $W$ boson mass from the ratio
of $W$ and $Z$ total cross sections \cite{Brock:1999ep,Rajagopalan:1996wx,Giele:1998uh,Shpakov:2000eq}. 

The PDF uncertainty on $r$ is reduced ($\Delta r/r\approx\left|\Delta X/X-\Delta Y/Y\right|$)
if $X$ and $Y$ are strongly correlated, cf. Eq.~(\ref{dfrat}).
It is enhanced ($\Delta r/r\approx\Delta X/X+\Delta Y/Y$) if $X$
and $Y$ are strongly anticorrelated. It is therefore beneficial to
construct the cross section ratios from pairs of correlated cross
sections to reduce the PDF uncertainties. We will examine a ratio
between the highly-correlated $W$ and $Z$ total cross sections in
Section~\ref{sub:WZStrange}. We will also explore the inclusive
$t\bar{t}$ production cross section, potentially useful for building
ratios when an anti-correlation with $W$ and $Z$ cross sections
occurs. $W$, $Z$, $t\bar t$ total cross section values
for 45 CTEQ6.6 sets suitable for the computation of various PDF correlations 
can be downloaded at \cite{CTEQ66WebsitePDFs}.

Leading channels for neutral Higgs boson ($h^{0}$) production in
the standard model \cite{Spira:1995mt} are also investigated, including
gluon-gluon fusion $gg\rightarrow h^{0},$ the dominant Higgs production
mechanism; vector boson fusion $q\bar{q}\rightarrow WW\rightarrow h^{0}$
(or $q\bar{q}\rightarrow ZZ\rightarrow h^{0}$), a prominent channel
employed at the LHC for both discovery and measurement of $h^{0}WW$
(or $h^{0}ZZ$) couplings; and associated production of Higgs boson
with a massive weak boson, $q\bar{q}\rightarrow Wh^{0}$ ($q\bar{q}\rightarrow Zh^{0}$),
the leading discovery mode for Higgs masses $m_{h}<200$ GeV at the
Tevatron. 

In addition, Higgs production channels in MSSM are considered: neutral
CP-even or CP-odd Higgs boson production via bottom-quark annihilation,
$b\bar{b}\rightarrow h$ ($h=h^{0},H^{0},A^{0}$) \cite{Spira:1997dg};
charged Higgs boson production via $s,$ $c,$ $b$ scattering, $c\bar{s}+c\bar{b}\rightarrow h^{+}$
\cite{DiazCruz:2001gf}; and associated production $q\bar{q}^{\prime}\rightarrow W\rightarrow Ah^{\pm}$
of CP-odd ($A$) and charged Higgs bosons \cite{Kanemura:2001hz,Cao:2003tr},
with Higgs masses related as $m_{h^{\pm}}^{2}=m_{A}^{2}+M_{W}^{2}$
at the Born level\textbf{.} The $b\bar{b}\rightarrow h$ and $c\bar{s}+c\bar{b}\rightarrow h^{+}$
cross sections may be greatly enhanced if the ratio $\tan\beta=v_{u}/v_{d}$
of vacuum expectation values for up- and down-type Higgs doublets
is of order 10 or more, a possibility that remains compatible with
LEP, Tevatron, and other constraints \cite{:2001xx}. To be specific,
we evaluate $q\bar{q}h$ couplings at the tree level and choose effective
$\tan\beta=50$; but most of our results (presented as cross section
ratios for different PDF sets and correlation cosines) are independent
of the $\tan\beta$ value. 

Supersymmetric neutral Higgs production $b\bar{b}\rightarrow h$ is
sensitive largely to $b(x,\mu)$ and $g(x,\mu)$, while charged Higgs
production probes a combination of $s(x,\mu),$ $c(x,\mu),$ and $b(x,\mu)$.
Long-distance hadronic functions entering these processes require
a general-mass scheme approach \cite{Lai:2007dq,Pumplin:2007wg,Berge:2005rv,Belyaev:2005bs}.
In hard-scattering inclusive cross sections, heavy-flavor mass corrections
enter through ratios $m_{c,b}^{2}/p_{i}^{2}$ and are suppressed at
large momentum scales $p_{i}^{2}\gg m_{c,b}^{2}$. We therefore neglect
the charm and bottom quark masses in hard-scattering matrix elements
at TeV energies, while keeping the mass effects inside the PDFs.

\subsection{Total cross sections at the Tevatron and the LHC\label{sub:TotalXSecsTevLHC}}

\begin{figure}
\noindent \begin{centering}\includegraphics[width=0.7\textwidth,keepaspectratio]{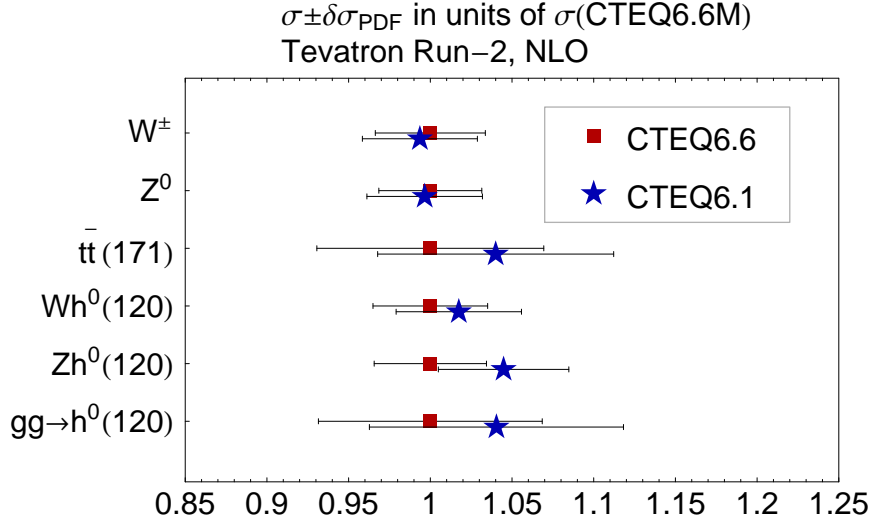}\\
 (a)\par\end{centering}

\noindent \begin{centering}\includegraphics[width=0.7\textwidth,keepaspectratio]{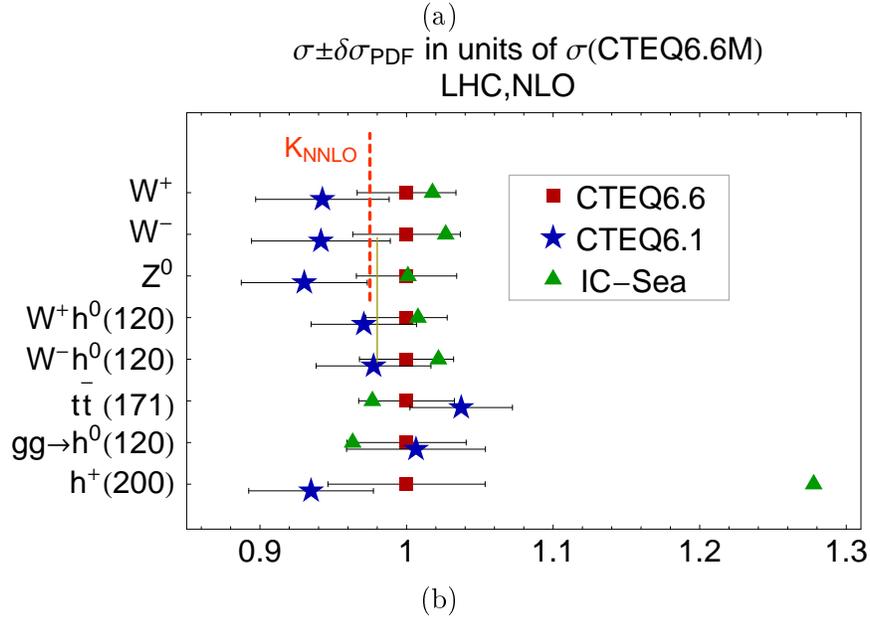}\\
(b)\par\end{centering}

\caption{Representative CTEQ6.6 (red boxes) and 6.1 (blue stars) total cross
sections and PDF uncertainties at the Tevatron and LHC, normalized
to the CTEQ6.6M cross section. Green triangles indicate CTEQ6.6 cross
sections obtained under an assumption of a strong sea-like intrinsic
charm production (IC-Sea-3.5\%). \label{fig:sigmatot} }
\end{figure}

To discuss common differences between the general-mass (CTEQ6.6) and
zero-mass (CTEQ6.1) predictions, we compute several total cross sections
sensitive to light-quark, gluon, or heavy-quark scattering using the
NLO programs WTTOT \cite{WTTOT} and MCFM \cite{Campbell:2004ch,Campbell:2005bb,Campbell:2000bg}.
Figure~\ref{fig:sigmatot} shows several CTEQ6.6 and CTEQ6.1 total
cross sections and their PDF uncertainties at the $p\bar{p}$ collider
Tevatron ($\sqrt{s}=1.96\mbox{ TeV}$) and the $pp$ Large Hadron
Collider ($\sqrt{s}=14\mbox{ TeV}$). They are plotted as ratios to
CTEQ6.6M predictions. We do not show predictions for the CTEQ6.5 set,
as those agree with CTEQ6.6 within the uncertainties. 

The processes shown in the figure for the Tevatron occur at relatively
large momentum fractions $x,$ where CTEQ6.6 and CTEQ6.1 PDFs agree
well. Consequently the CTEQ6.6 and CTEQ6.1 cross sections for the
Tevatron processes coincide within the PDF uncertainties, as illustrated
in Fig.~\ref{fig:sigmatot}(a). The magnitudes of the PDF uncertainties
also remain about the same. 

At the LHC {[}Fig.~\ref{fig:sigmatot}(b)], CTEQ6.6 cross sections
for $W$ and $Z$ boson production are enhanced by $6-7\%$ compared
to CTEQ6.1 because of the larger magnitudes of $u\!\!\!{}^{^{(-)}}(x,Q)$
and $d\!\!\!{}^{^{(-)}}(x,Q)$ in the relevant range of $x\sim M_{W,Z}/\sqrt{s}=10^{-3}-10^{-2}$.
The CTEQ6.6 light-quark parton luminosities ${\cal L}_{q_{i}\bar{q_{j}}}(x_{1},x_{2},Q)=q_{i}(x_{1},Q)\,\bar{q}_{j}(x_{2},Q)$
(and therefore CTEQ6.6 cross sections) are larger at such $x$ by
$2\,\delta f\approx6\%$, where $\delta f\approx3\%$ is the typical
increase in the GM light-quark PDFs compared to the ZM PDFs. The Hessian
PDF error obtained by our standard $90\%$ c.l. criterion has decreased
from $4.5-5\%$ in CTEQ6.1 to 3.5\% in CTEQ6.6, mostly because new
DIS experimental data were included in the CTEQ6.6 fit. The differences
between CTEQ6.6 and CTEQ6.1 exceed the magnitude of the NNLO correction
to $W$ and $Z$ cross sections of order $2\%$ \cite{Hamberg:1990np,Harlander:2002wh,Anastasiou:2003ds,Anastasiou:2003yy},
indicated by a dashed line. Uncertainties of this size have important
implications for the calibration of the LHC luminosity.

Other cross sections dominated by light (anti)quark scattering at
$x\sim10^{-2}$ increase by comparable amounts. For example, the CTEQ6.6
cross sections for associated $W^{\pm}h$ or $Z^{0}h$ boson production
exceed CTEQ6.1 cross sections by $3-4\%$. In contrast, in processes
dominated by gluon or heavy-quark scattering, such as $t\bar{t}$
production or $gg\rightarrow h^{0}X$, the general tendency for the
CTEQ6.6 cross sections is to be a few percent lower compared to CTEQ6.1.
The CTEQ6.6M $c\bar{s}+c\bar{b}\rightarrow h^{+}$ cross section is
enhanced with respect to CTEQ6.1M by its larger strangeness PDF (cf.
Fig.~\ref{fig:CTEQ66strange}), despite some suppression of the $c\bar{b}$
contribution to this process. The LHC cross sections in Fig.~\ref{fig:sigmatot}(b)
may change substantially if a fraction of charm quarks is produced
through the nonperturbative {}``intrinsic'' mechanism. For example,
the rate for production $sc+sb\rightarrow h^{\pm}$ of MSSM charged
Higgs bosons with mass 200 GeV would increase by 30\% if sea-like
intrinsic charm contributions carry 3.5\% of the parent nucleon's
momentum, the maximal amount tolerated in the fit.

\subsection{$W$ and $Z$ boson production cross sections \label{sub:W-and-Z}}

\subsubsection{CTEQ6.6 vs. other PDF sets}

Large groups of collider cross sections, notably those dominated largely
by quark scattering or largely by gluon scattering, exhibit correlated
dependence on PDFs. Production of charged $W^{\pm}$ and neutral $Z^{0}$
bosons are essential quark-quark scattering processes that show such
correlations.

\begin{figure}
\noindent \begin{centering}\includegraphics[width=0.5\columnwidth,keepaspectratio]{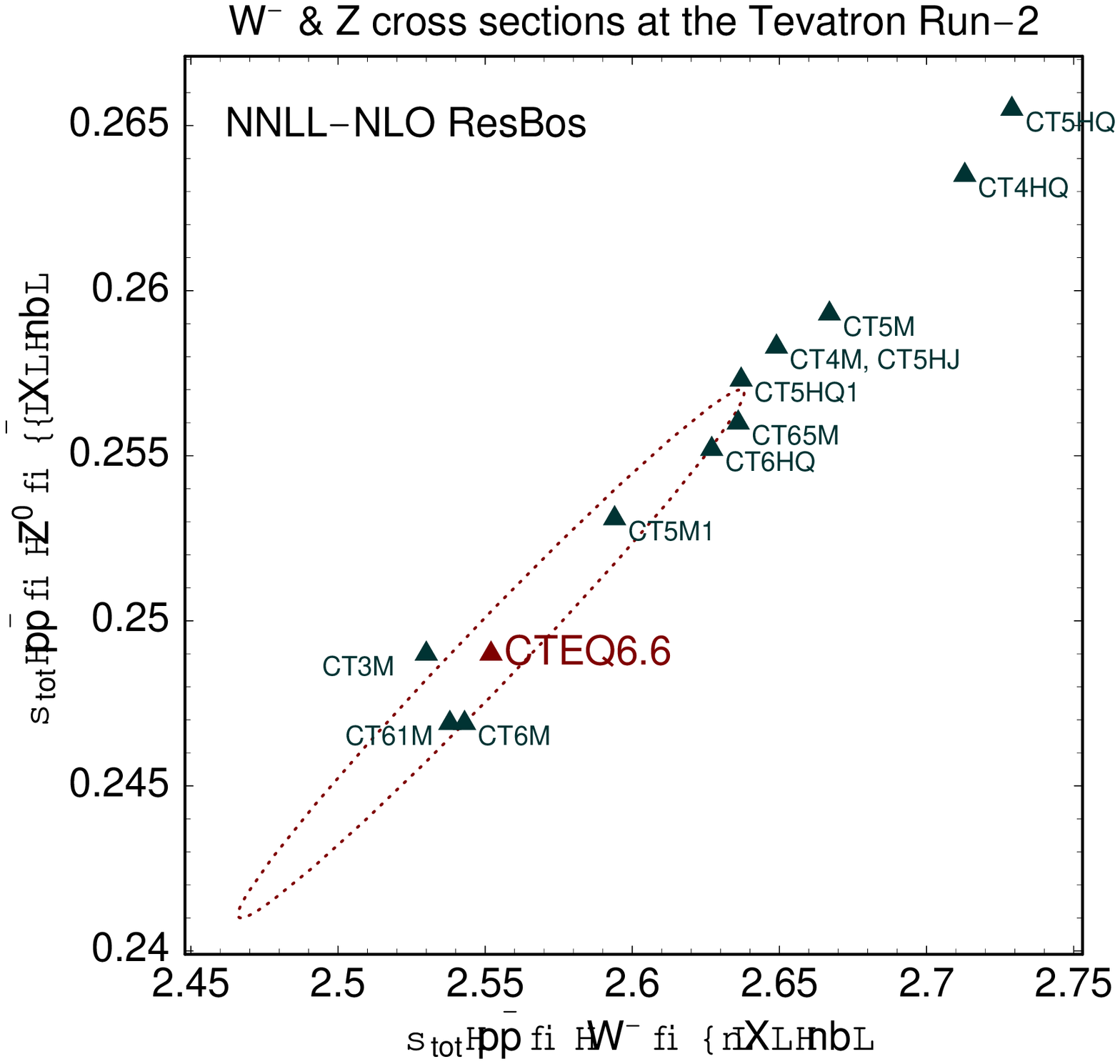}\includegraphics[width=0.5\columnwidth,keepaspectratio]{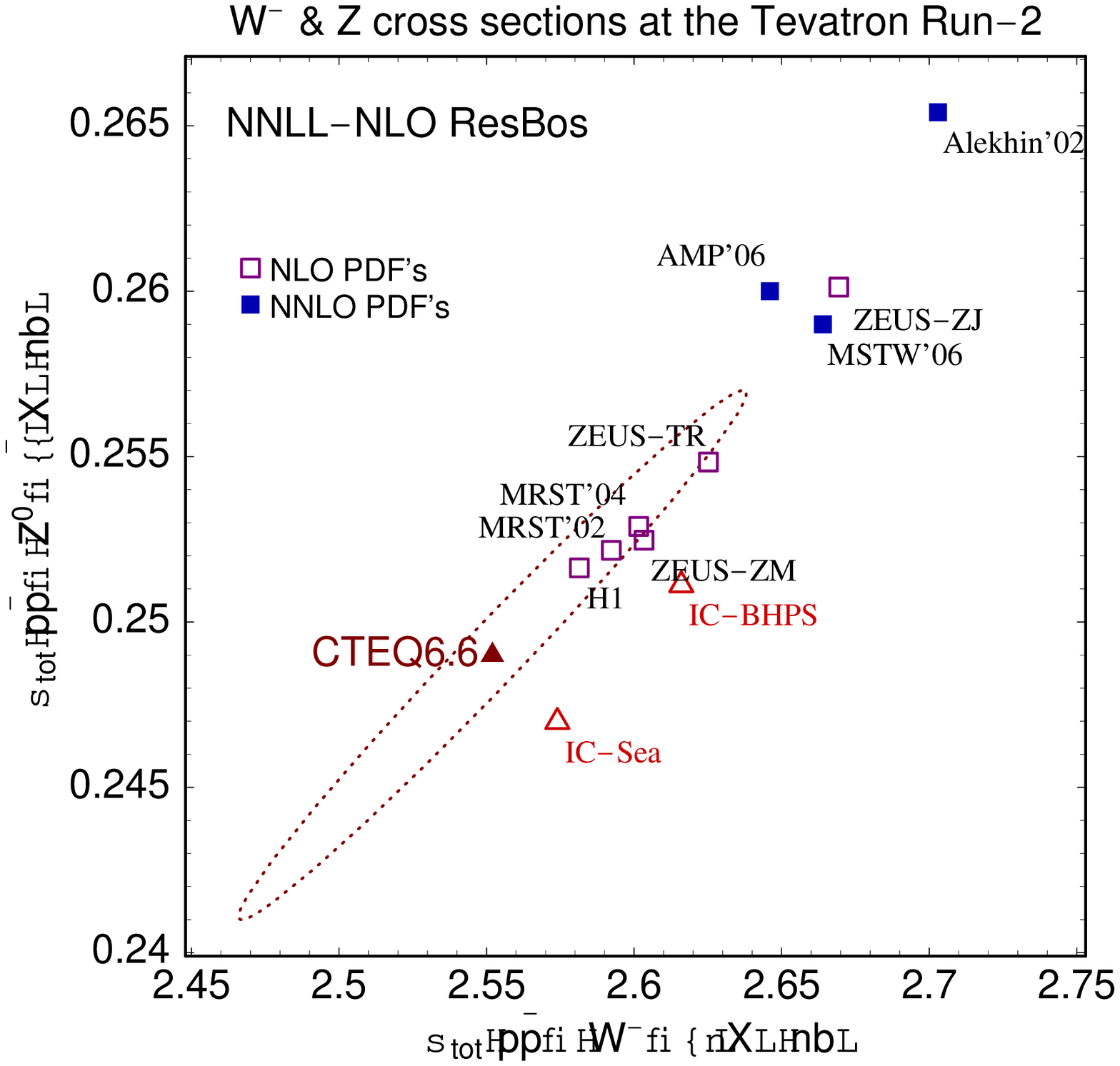}\\
(a)\hspace{3in}(b)\\
\par\end{centering}

\caption{CTEQ6.6 NNLL-NLO $W^{\pm}$ and $Z$ production cross sections in
the Tevatron Run-2, compared with predictions for other PDF sets by
(a) CTEQ and (b) other groups. \label{fig:WZXsecsTev2} }
\end{figure}

\begin{figure}
\noindent \begin{centering}\includegraphics[width=0.5\columnwidth,keepaspectratio]{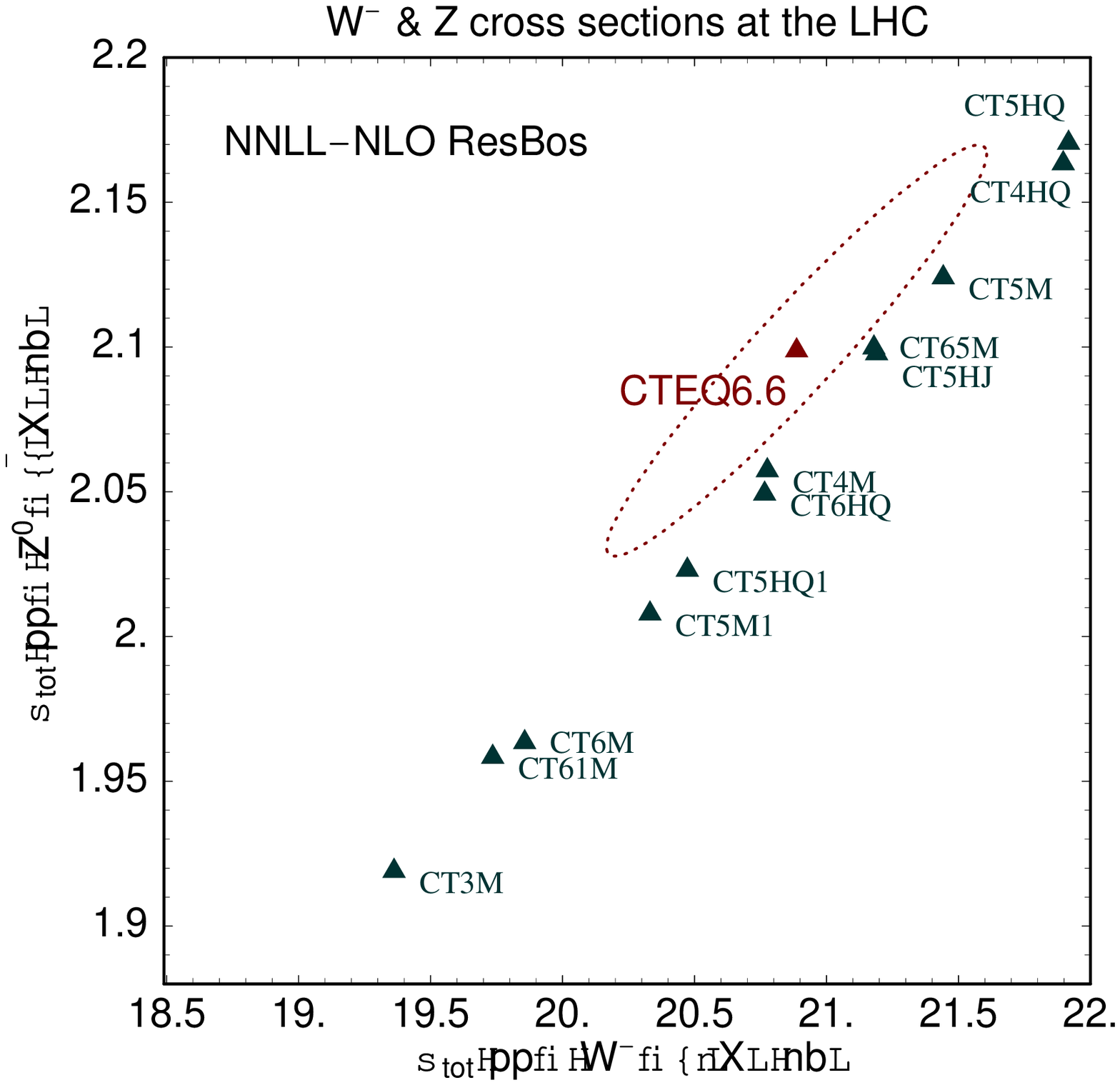}\includegraphics[width=0.5\columnwidth,keepaspectratio]{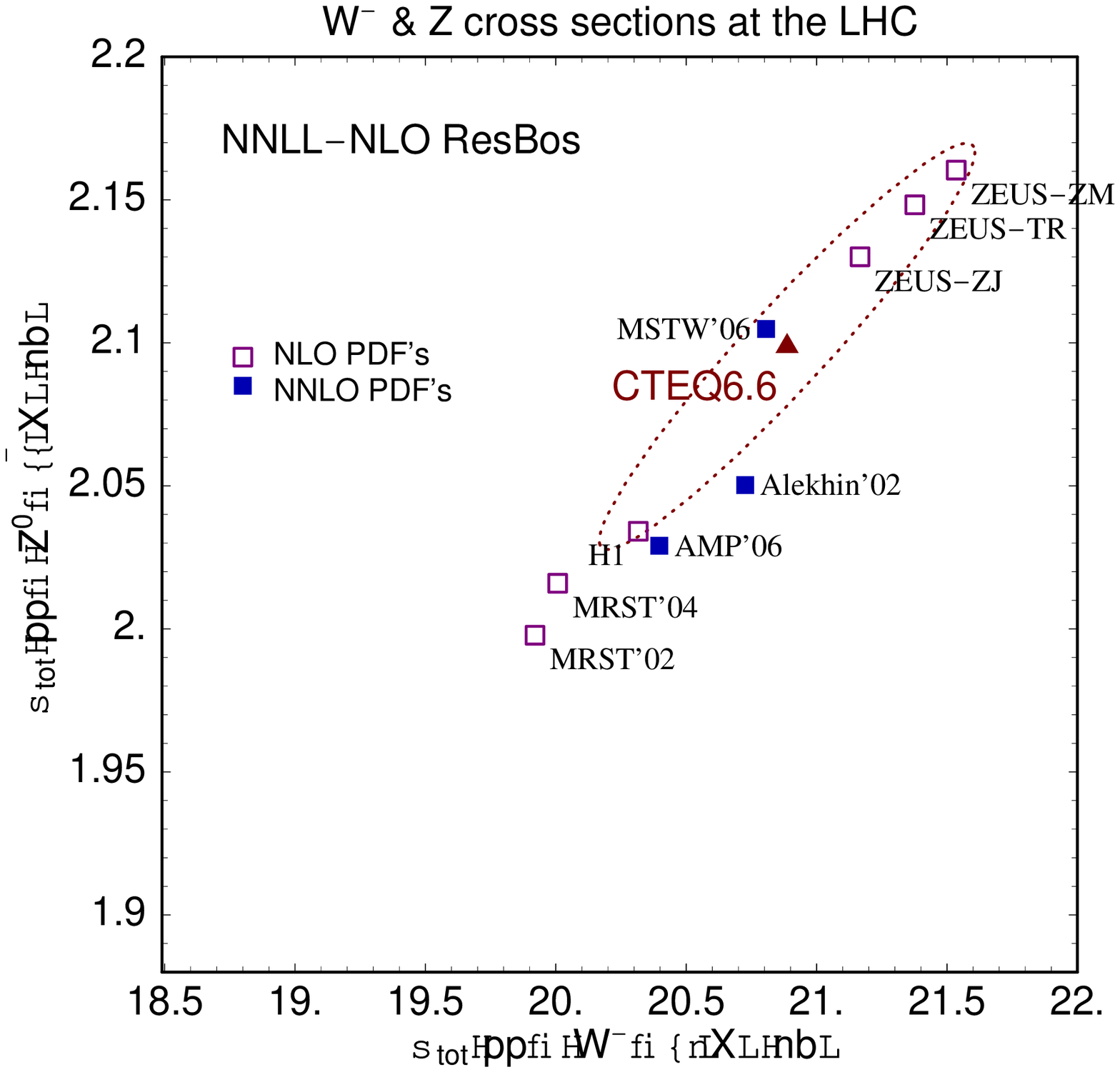}\\
(a)\hspace{3in}(b)\par\end{centering}

\caption{CTEQ6.6 NNLL-NLO $W^{\pm}$ and $Z$ production cross sections in
the LHC, compared with predictions for (a) different PDF sets by CTEQ
and (b) other groups. \label{fig:WZXsecsLHC}}
\end{figure}

Figs.~\ref{fig:WZXsecsTev2} and \ref{fig:WZXsecsLHC} summarize
predictions for the total cross sections $\sigma$ of $W$ and $Z$
production, obtained in the NNLL-NLO resummation calculation (of order
${\cal O}(\alpha_{s})+$leading higher-order logarithms) \cite{Ladinsky:1994zn,Balazs:1997xd,Landry:2002ix}
and using recent PDF sets by CTEQ \cite{Lai:1994bb,Lai:1996mg,Lai:1997vu,Lai:1999wy,Pumplin:2002vw,Stump:2003yu,Kretzer:2003it,Tung:2006tb},
Alekhin/AMP \cite{Alekhin:2002fv,Alekhin:2006zm}, H1 \cite{Adloff:2003uh},
MRST/MSTW \cite{Martin:2002aw,Martin:2003sk,Martin:2004ir,Martin:2007bv},
and ZEUS \cite{Chekanov:2002pv,Chekanov:2005nn} groups. We include
decays of the massive bosons into lepton pairs in the improved Born
approximation \cite{Balazs:1997xd,Nadolsky:2003fz}. To provide a
visual measure of the PDF uncertainty, each figure shows an error
ellipse corresponding to our usual tolerance criterion.%
\footnote{In two-dimensional plots, this criterion corresponds to probability
somewhat smaller than 90\%. %
} The ellipses are found from Eqs.~(\ref{ellipse1}) and (\ref{ellipse2}),
using the parameters listed in Table~\ref{tab:CorrsStdCandles}.

At the Tevatron, the CTEQ6.6 cross sections lie close to the CTEQ6.1
and CTEQ6.5 predictions {[}Fig.~\ref{fig:WZXsecsTev2}(a)], i.e.,
the dependence on the heavy-flavor scheme is relatively weak in this
case. The CTEQ6.6 cross sections agree well with predictions based
on the PDF sets by the other groups {[}Fig.~\ref{fig:WZXsecsTev2}(b)]. 

At the LHC, the general-mass CTEQ6.6 cross sections exceed the zero-mass
CTEQ6.1 cross sections by $6-7\%$, as shown in Fig.~\ref{fig:WZXsecsLHC}(a).
The CTEQ6.6 and 6.5 $Z^{0}$ cross sections are about the same. The
CTEQ6.6 $W^{\pm}$ cross section is somewhat smaller than the CTEQ6.5
cross section and does not lie on the same line in the $W-Z$ plane
as the previous CTEQ sets. The predictions based on the latest CTEQ6.6,
MSTW'06, and AMP'06 PDFs agree within 3\% {[}Fig.~\ref{fig:WZXsecsLHC}(b)]. 

The total cross sections shown here are somewhat affected by higher-order
contributions, not included under the current NNLL-NLO approximation.
A prediction of \emph{absolute} magnitudes of $W$ and $Z$ cross
sections with accuracy 1\% would require to simultaneously evaluate
NNLO QCD contributions of order ${\cal O}(\alpha_{s}^{2})$ \cite{Hamberg:1990np,Harlander:2002wh,Anastasiou:2003ds,Anastasiou:2003yy}
and NLO electroweak contributions of order ${\cal O}(\alpha_{s}\alpha_{EW})$
\cite{Baur:2001ze,Baur:1998kt,Arbuzov:2007db,Kuhn:2007cv,Kuhn:2007qc,CarloniCalame:2004qw,CarloniCalame:2006zq,Calame:2007cd,Gerber:2007xk}
for both the hard cross sections and PDFs, including all relevant
spin correlations \cite{Cao:2004yy,Frixione:2004us}. This level of
accuracy is not yet achieved. However, the higher-order terms rescale
the NLO hard cross sections by overall factors with weak dependence
on the PDFs \cite{Anastasiou:2003ds,Anastasiou:2003yy}. Therefore
our NLO-NNLL total cross sections should reasonably estimate true
\emph{relative} differences caused by the PDFs. 

The NNLL-NLO cross sections for AMP'2006 and MSTW'2006 are computed
using the NNLO PDFs, since the NLO PDFs for these sets are not available.
Such combination is acceptable at the order we are working. For example,
mixing of NLO-NNLL and NNLO orders is of little consequence in the
case of the MRST'2004 set, for which the replacement of the NLO PDFs
by the NNLO PDFs changes the cross sections by about 1.5\%. Variations
of this magnitude are clearly permissible until a more precise computation
is fully developed.

\subsubsection{Correlations between $W,Z$ cross sections and PDFs \label{sub:CorrWZatLHCandPDFs}}

Although strong PDF-induced correlations between the $W$ and $Z$
cross sections are observed at both colliders, the mechanism driving
these correlations at the LHC is not the same as at the Tevatron.
The essential point is that, although the weak bosons are mostly produced
in $u$ and $d$ quark-antiquark scattering, this dominant process
may contribute little to the PDF uncertainty because of tight constraints
imposed on the up- and down-quark PDFs by the DIS and Drell-Yan data. 

Instead, a substantial fraction of the PDF uncertainty at the LHC
(but not at the Tevatron) is contributed by sizable, yet less constrained,
contributions from heavy-quark ($s,$ $c,$ $b$) scattering. Subprocesses
with initial-state $s,$ $c,$ and, to a smaller extent, $b$ and
$g$ partons deliver up to 20\% of the NLO rate at the LHC, compared
to 2-4\% at the Tevatron. All these partons are correlated with the
gluons via DGLAP evolution, so that the LHC $W$ and $Z$ cross sections
are particularly sensitive to the uncertainty in the gluon PDF. Consequently,
$W$ and $Z$ cross sections at the LHC are better correlated with
processes dominated by gluon scattering in the comparable kinematical
range and not necessarily with $u$ and $d$ quark scattering, in
striking contrast to the Tevatron. 

\begin{figure}
\begin{centering}\includegraphics[width=0.5\columnwidth,height=8cm,keepaspectratio]{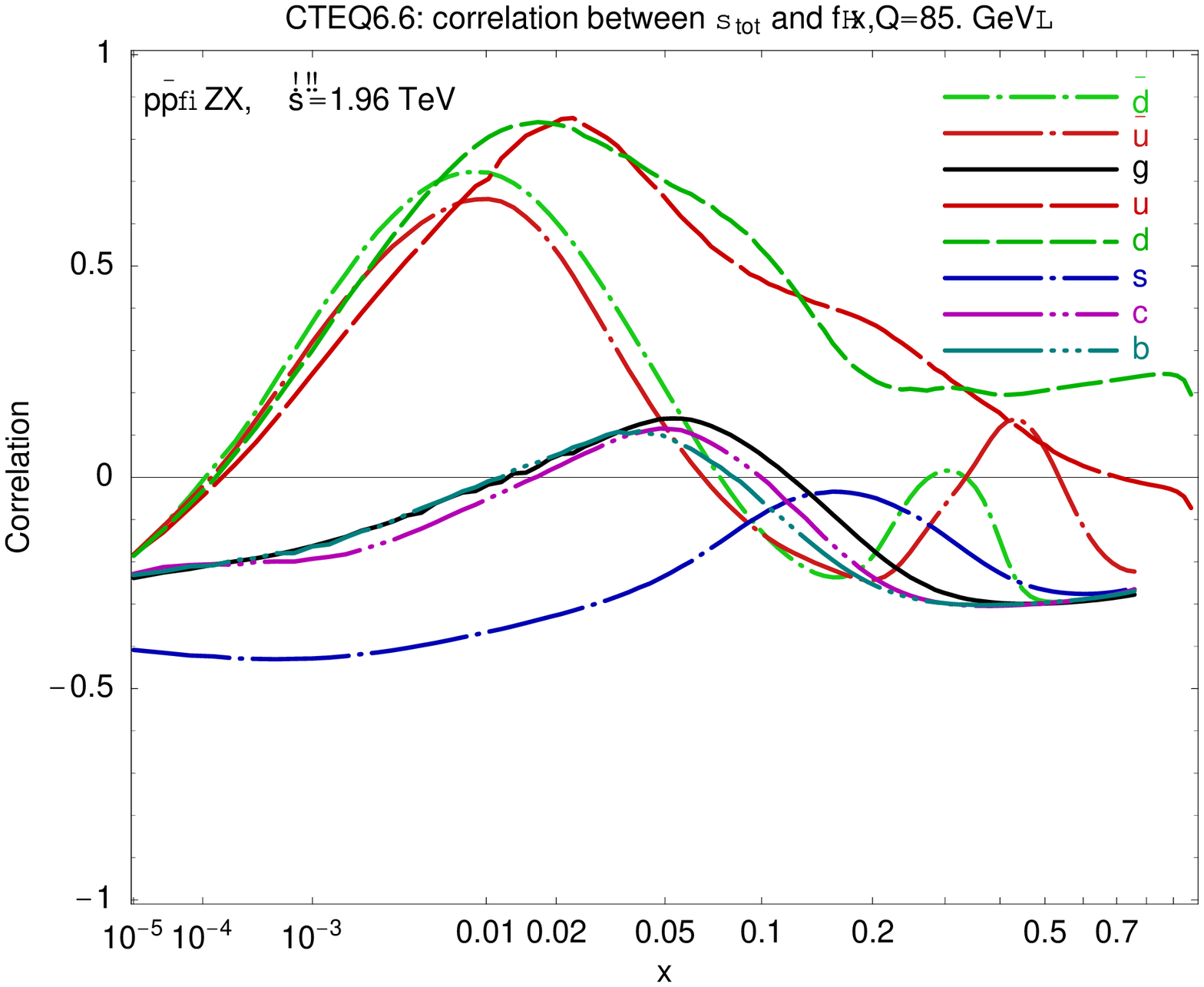}\includegraphics[width=0.5\columnwidth,height=8cm,keepaspectratio]{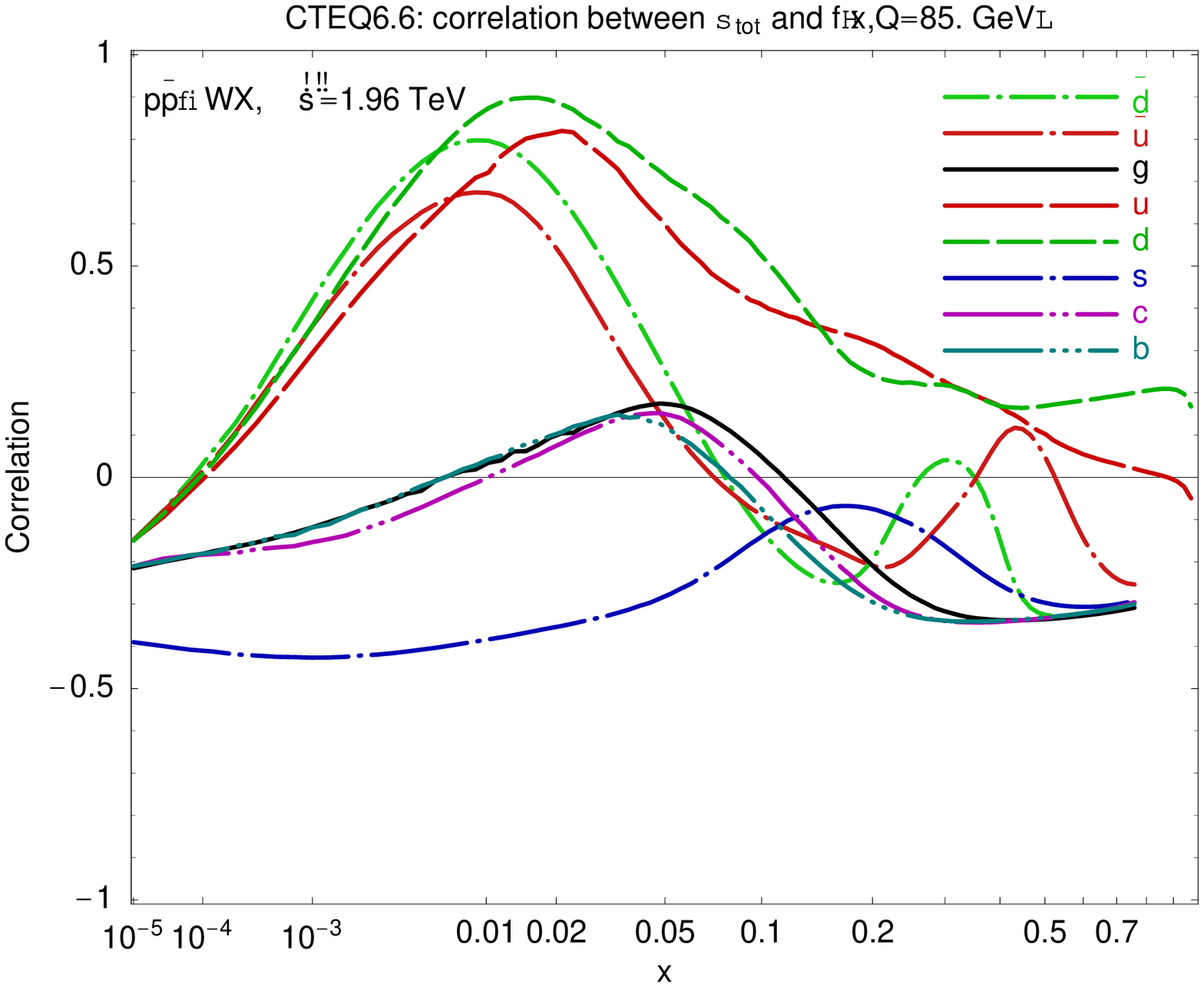}\par\end{centering}

\begin{centering}(a)\hspace{3in}(b)\par\end{centering}

\begin{centering}\includegraphics[width=0.5\columnwidth,height=8cm,keepaspectratio]{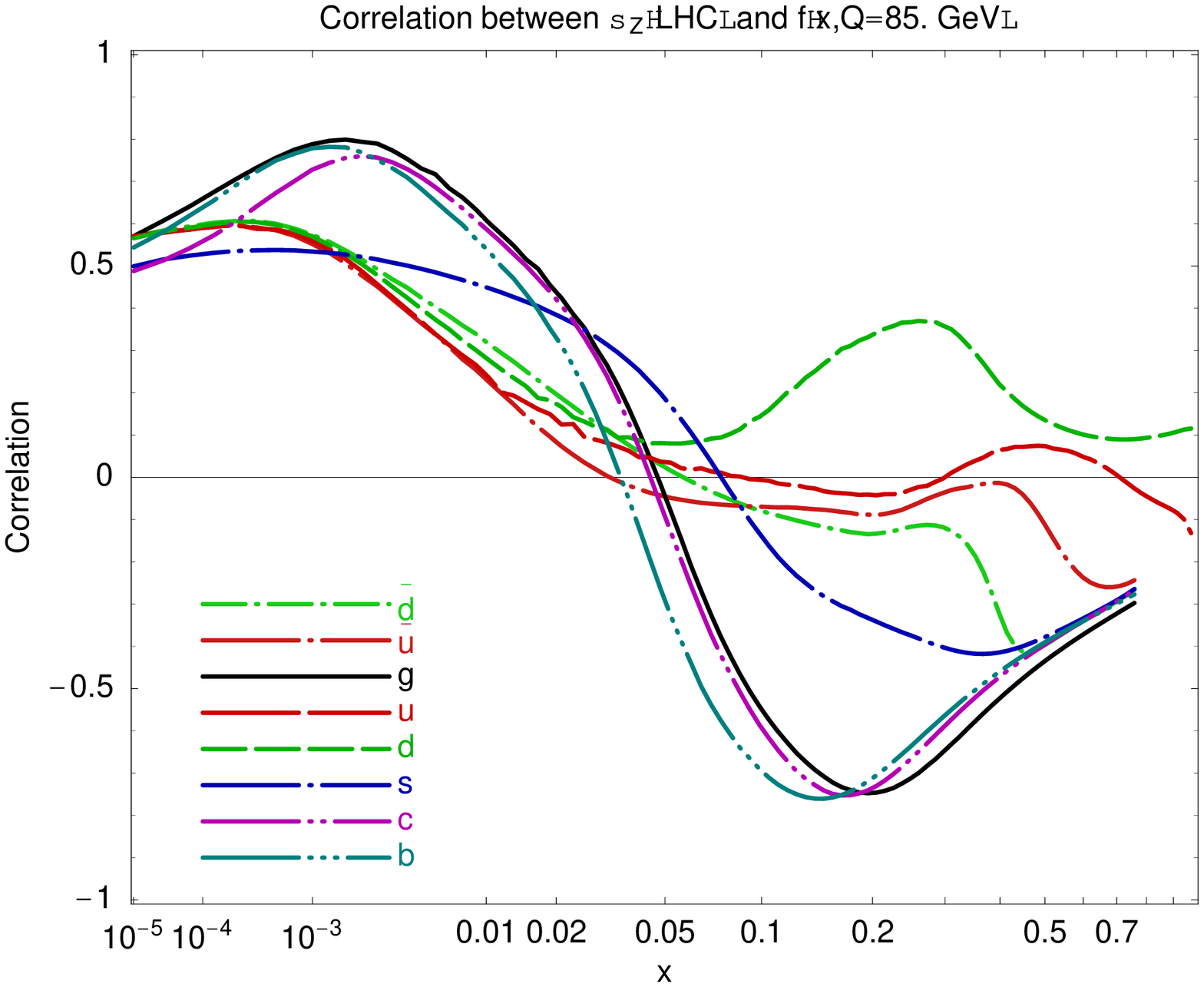}\includegraphics[width=0.5\columnwidth,height=8cm,keepaspectratio]{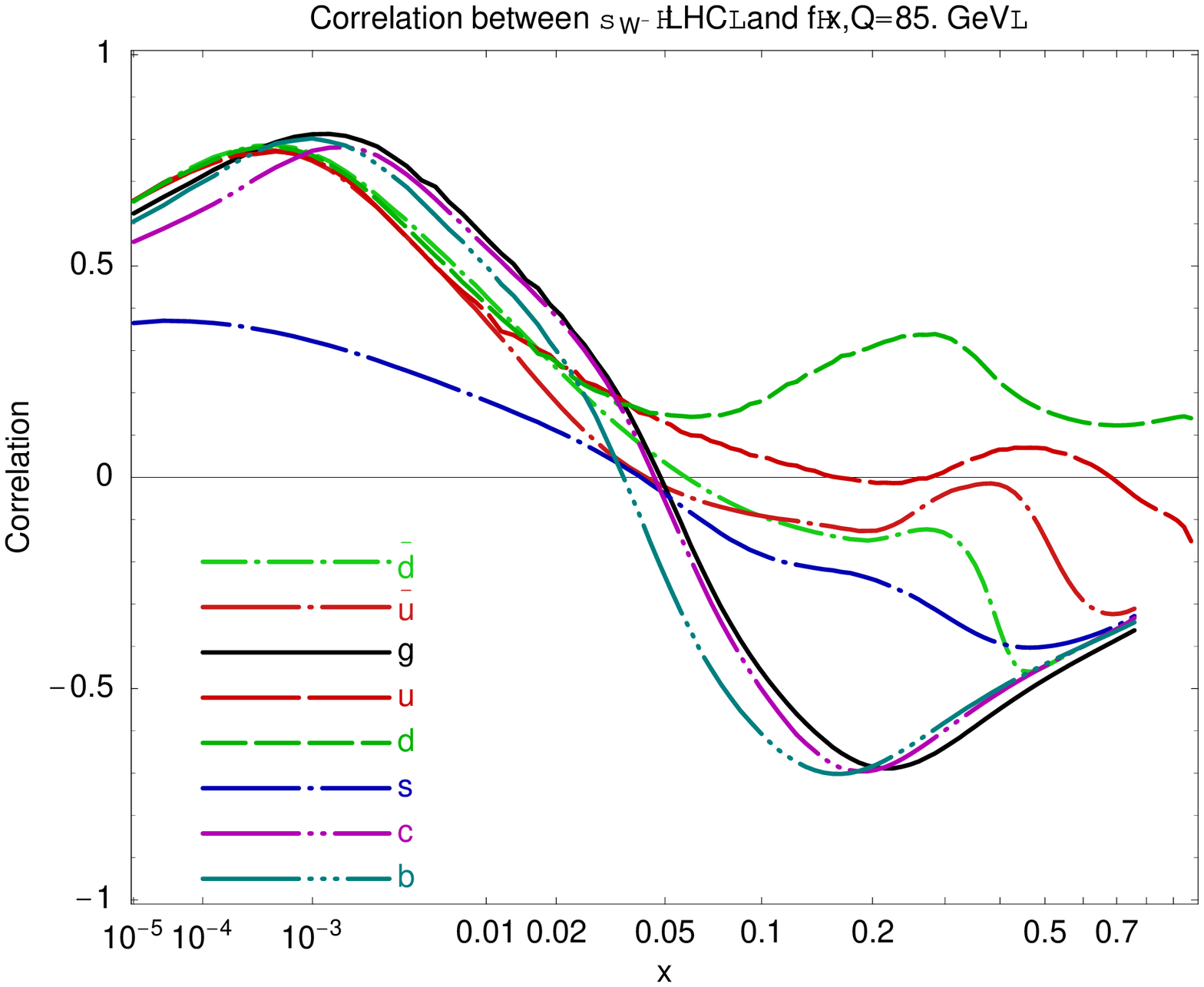}\par\end{centering}

\begin{centering}(c)\hspace{3in}(d)\par\end{centering}

\caption{(a,b) Correlation cosine ($\cos\varphi$) between the total cross
sections for $Z^{0}$ and $W^{\pm}$ production at the Tevatron and
PDFs, plotted as a function of $x$ for  $Q=85$ GeV; (c,d) the same
for the LHC. \label{fig:WZCorr}}
\end{figure}

To illustrate this point, Fig.~\ref{fig:WZCorr} shows correlation
cosines ($\cos\varphi$) between the $W,$ $Z$ cross sections and
PDFs $f_{a}(x,Q)$ of different flavors, evaluated as functions of
the momentum fraction $x$ at $Q=85$ GeV. The largest correlations
between the cross section and PDFs occur at momentum fractions $x$
of order $M_{W,Z}/\sqrt{s}$ corresponding to central rapidity production,
i.e., at $x\sim0.04$ at the Tevatron and 0.006 at the LHC. PDF flavors
with a very large correlation are associated with the major part of
the PDF uncertainty in the physical cross section. Additional constraints
on this flavor would help reduce the PDF uncertainty. 

At the Tevatron Run-2 {[}Figs.~\ref{fig:WZCorr}(a) and \ref{fig:WZCorr}(b)],
large correlations exist with $u$, $\bar{u},$ $d,$ and $\bar{d}$
PDFs, with $\cos\varphi$ reaching 0.95. No tangible correlation occurs
with PDFs for $s,$ $c,$ $b$ (anti)quarks and gluons. 

At the LHC {[}Figs.~\ref{fig:WZCorr}(c) and \ref{fig:WZCorr}(d)],
the largest correlations are driven by charm, bottom, and gluon PDFs,
followed by smaller correlations with $u,$ $d,$ and $s$ quarks.
The correlation with the $u$ and $d$ PDFs is reduced, although not
entirely eliminated. A large correlation with the gluon at $x\sim0.005$
is accompanied by a large anti-correlation ($\cos\varphi\sim-0.8$)
with the gluon at $x\sim0.1-0.2,$ as a consequence of nucleon momentum
conservation (cf.~Section~\ref{sub:PDFCorrelations}). This feature
implies that the LHC $W,$ $Z$ cross sections are strongly anticorrelated
with new particle production in gluon or heavy-quark scattering in
the TeV mass range, and moderately anticorrelated with $t\bar{t}$
production (cf.~Section~\ref{sub:ttbCorr}). 

\subsubsection{Impact of strangeness \label{sub:WZStrange}}

We stress that the large correlation between the LHC $Z$ cross section
and the gluon distribution is generated primarily by the sizable $s\bar{s}$
scattering contribution, and not by a smaller $qg$ scattering contribution.
Similarly, a large correlation in $W$ boson production is driven
for the most part by the $c\bar{s}$ contribution. Yet the correlations
of $Z$ and $W$ cross sections with the strangeness PDF, taken on
its own, are rather small {[}see Figs.~\ref{fig:WZCorr}(c) and (d)].
Adjustments in $s(x,Q)$ can be easily compensated by other quark
PDFs without changing the physical $W$ and $Z$ cross sections. This
extra freedom is absent in the case of the gluon PDF, since it affects
all sea PDFs at once through perturbative evolution. 

\begin{figure}
\begin{centering}\includegraphics[width=0.5\columnwidth,height=8cm,keepaspectratio]{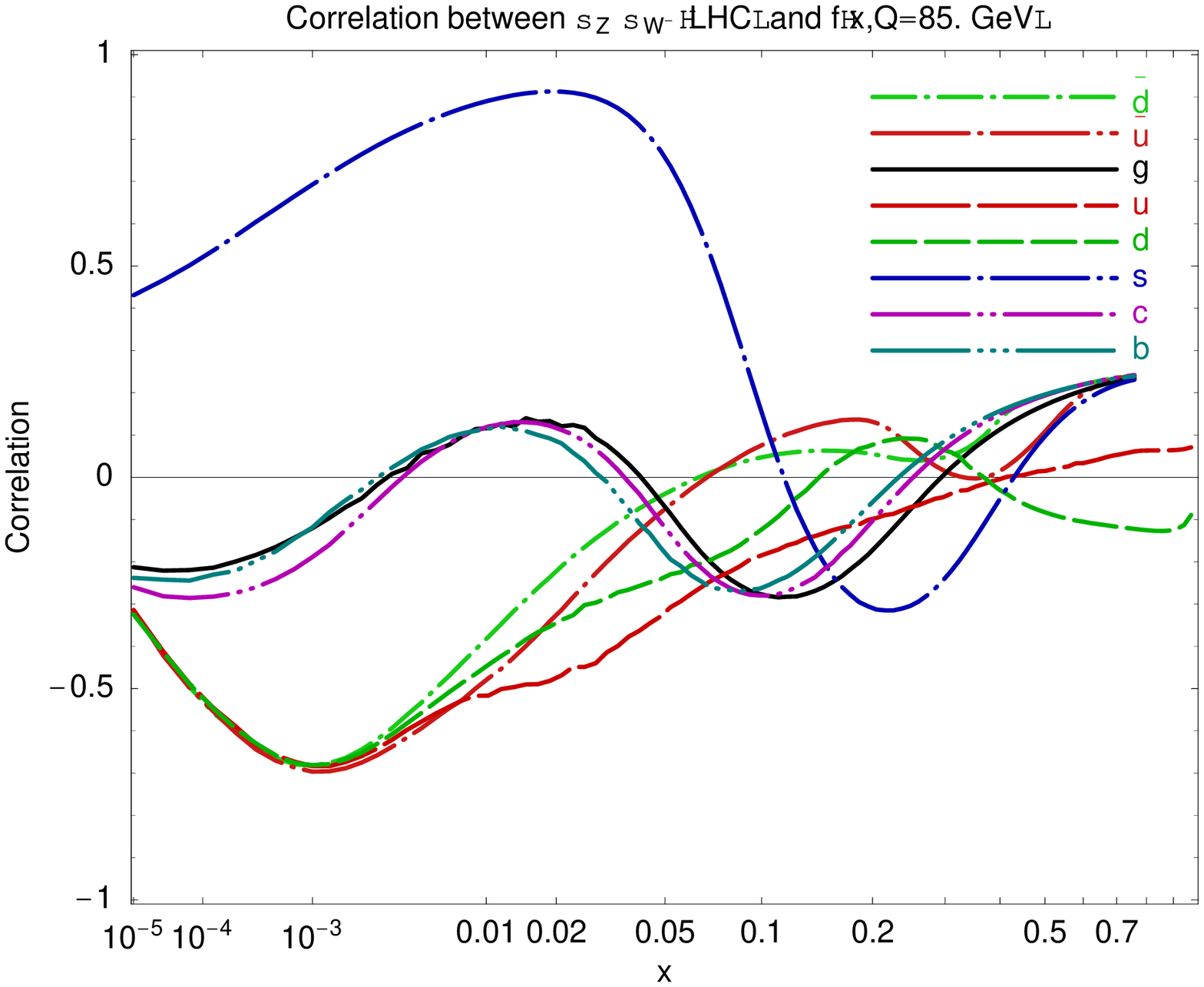}\par\end{centering}

\caption{Correlation cosine ($\cos\varphi$) between the ratio $\sigma_{Z}/\sigma_{W}$
of LHC total cross sections for $Z^{0}$ and $W^{\pm}$ production
at PDFs of various flavors, plotted as a function of $x$ for $Q=85$
GeV. \label{fig:ZWratCorr}}
\end{figure}

On the other hand, the ratio $r_{ZW}\equiv\sigma_{Z}/(\sigma_{W^{+}}+\sigma_{W^{-}})$
of the $Z^{0}$ and $W^{\pm}$ cross sections is very sensitive to
the uncertainty in strangeness. Nominally $r_{ZW}$ is an exemplary
{}``standard candle'' LHC observable because of the cancellation
of uncertainties inside the ratio. The CTEQ6.6 prediction $r_{ZW}=0.100\pm0.001$
is in an excellent agreement with the predictions based on the other
recent PDF sets. This result can be derived from Eq.~(\ref{dfrat})
by substituting the correlated parameters of the $W$ and $Z$ cross
sections in Table~\ref{tab:CorrsStdCandles}.

Fig.~\ref{fig:ZWratCorr} indicates that $r_{ZW}$ is mostly correlated
with the strangeness PDF $s(x,Q)$ in the region $0.01<x<0.05$. It is 
anticorrelated with $u\!\!\!{}^{^{(-)}}$ and
$d\!\!\!{}^{^{(-)}}$ at $x\sim10^{-3}$.  There is no tangible correlation
with the gluon, charm, and bottom PDFs. 

Since the strangeness is the least constrained distribution among
the light-quark flavors \cite{Lai:2007dq,Martin:2003sk,Olness:2003wz},
it is thus important to correctly model its uncertainty to estimate
$r_{ZW}.$ CTEQ6.6 is our first general-purpose PDF set that includes
an independent parametrization for strangeness. It predicts a larger
uncertainty in $s(x,Q)$ (hence, a larger $\Delta r_{ZW}$) than the
previous PDF analyses, which artificially linked $s(x,Q)$ to the
combination $\left(\bar{u}+\bar{d}\right)$ of the lightest sea quarks. 

The increased fractional uncertainty $\Delta r_{ZW}/r_{ZW}$ is related
to weaker correlation between the CTEQ6.6 $W$ and $Z$ cross sections
at the LHC. According to Eq.~(\ref{dfrat}), $\Delta r_{ZW}/r_{ZW}$
scales approximately as $(1-\cos\varphi)^{1/2}$, given that the $W$
and $Z$ fractional uncertainties are about the same ($\Delta\sigma_{W}/\sigma_{W}^{0}\approx\Delta\sigma_{Z}/\sigma_{Z}^{0}$).
The value of $\cos\varphi$ decreases from 0.998 in CTEQ6.1 to 0.956
in CTEQ6.6, or 0.982 if $s(x,Q)$ is fixed during the Hessian analysis
at its best-fit CTEQ6.6M shape. As a result of the smaller $\cos\varphi$,
$\Delta r_{ZW}/r_{ZW}$ increases threefold in the CTEQ6.6 prediction,
even though the fractional uncertainties $\Delta\sigma_{W,Z}/\sigma_{W,Z}$
are reduced.

\begin{figure}
\noindent \begin{centering}\includegraphics[width=0.5\columnwidth,height=50cm,keepaspectratio]{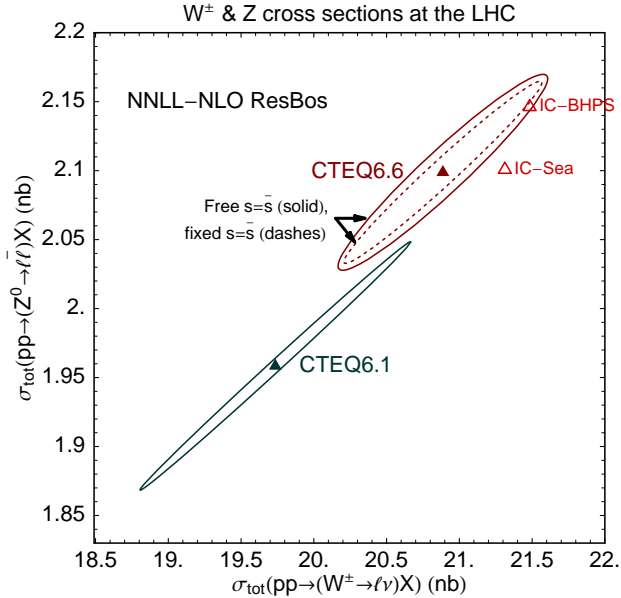}\par\end{centering}

\caption{$W$ and $Z$ correlation ellipses at the LHC obtained in the fits
with free and fixed strangeness, as well as with maximal intrinsic
charm contribution. \label{fig:WZLHCstrangeness}}
\end{figure}

The eccentricity of the $\sigma_{Z}-\sigma_{W}$ tolerance ellipse
grows with $\cos\varphi$, implying a narrow CTEQ6.1 ellipse and a
broader CTEQ6.6 ellipse shown in Fig.~\ref{fig:WZLHCstrangeness}.
The CTEQ6.6 ellipse narrows if the strangeness parameters are fixed
at their best-fit values as described above. Very different values
of $r_{ZW}$ can be obtained if one allows for the {}``intrinsic
charm'' contribution. In Fig.~\ref{fig:WZLHCstrangeness}, the empty
triangles correspond to the {}``maximal'' intrinsic charm scenarios.
These cross sections lie on the boundary or outside of the CTEQ6.6
tolerance ellipse. They can be potentially ruled out by measuring
$r_{ZW}$ precisely at the Tevatron, as can be deduced from Fig.~\ref{fig:WZXsecsTev2}(b).

\subsubsection{$W^{-}$ vs $W^{+}$ cross sections}

\begin{figure}
\noindent \begin{centering}\includegraphics[width=0.5\columnwidth,keepaspectratio]{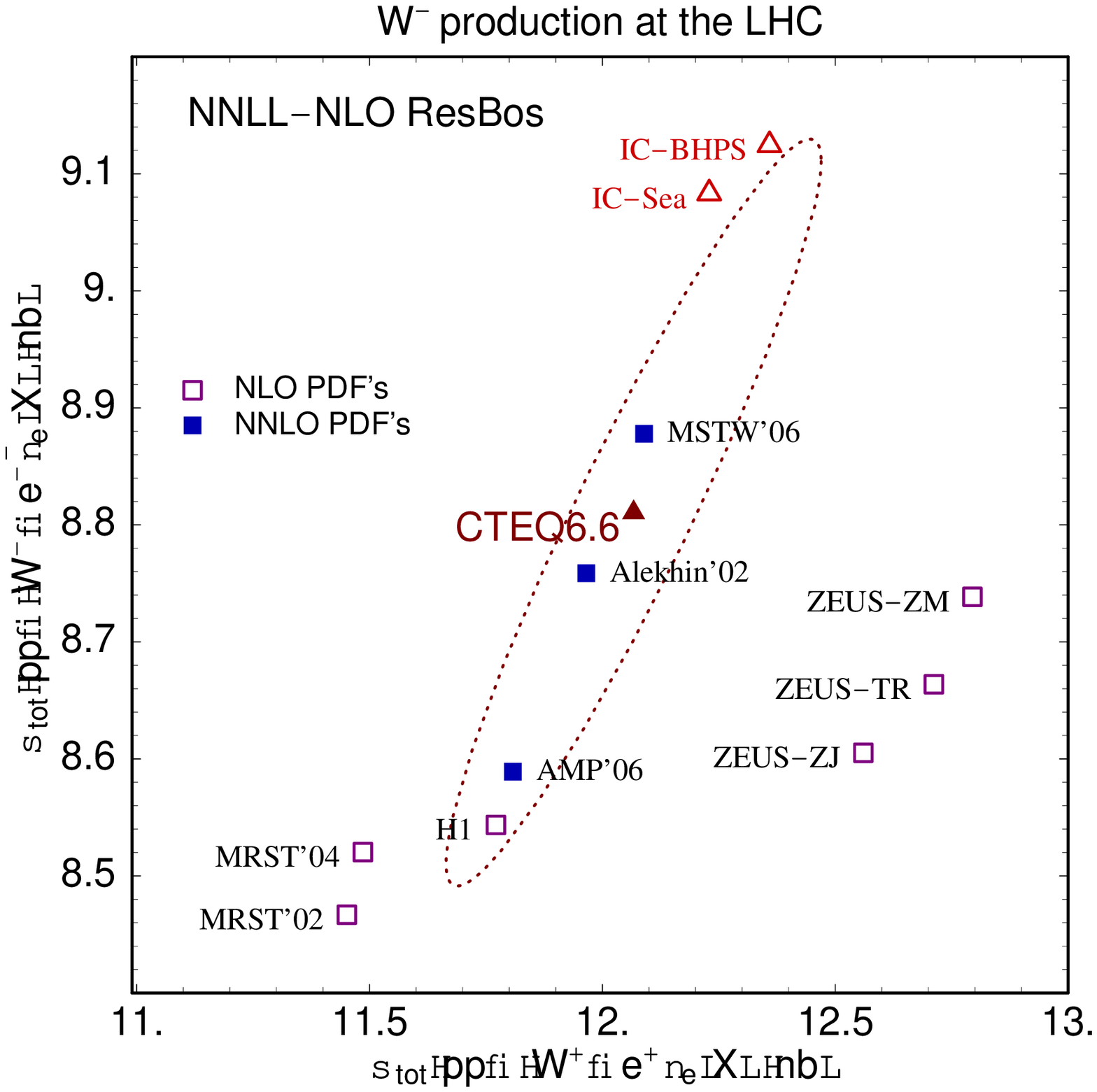}\includegraphics[width=0.5\columnwidth,keepaspectratio]{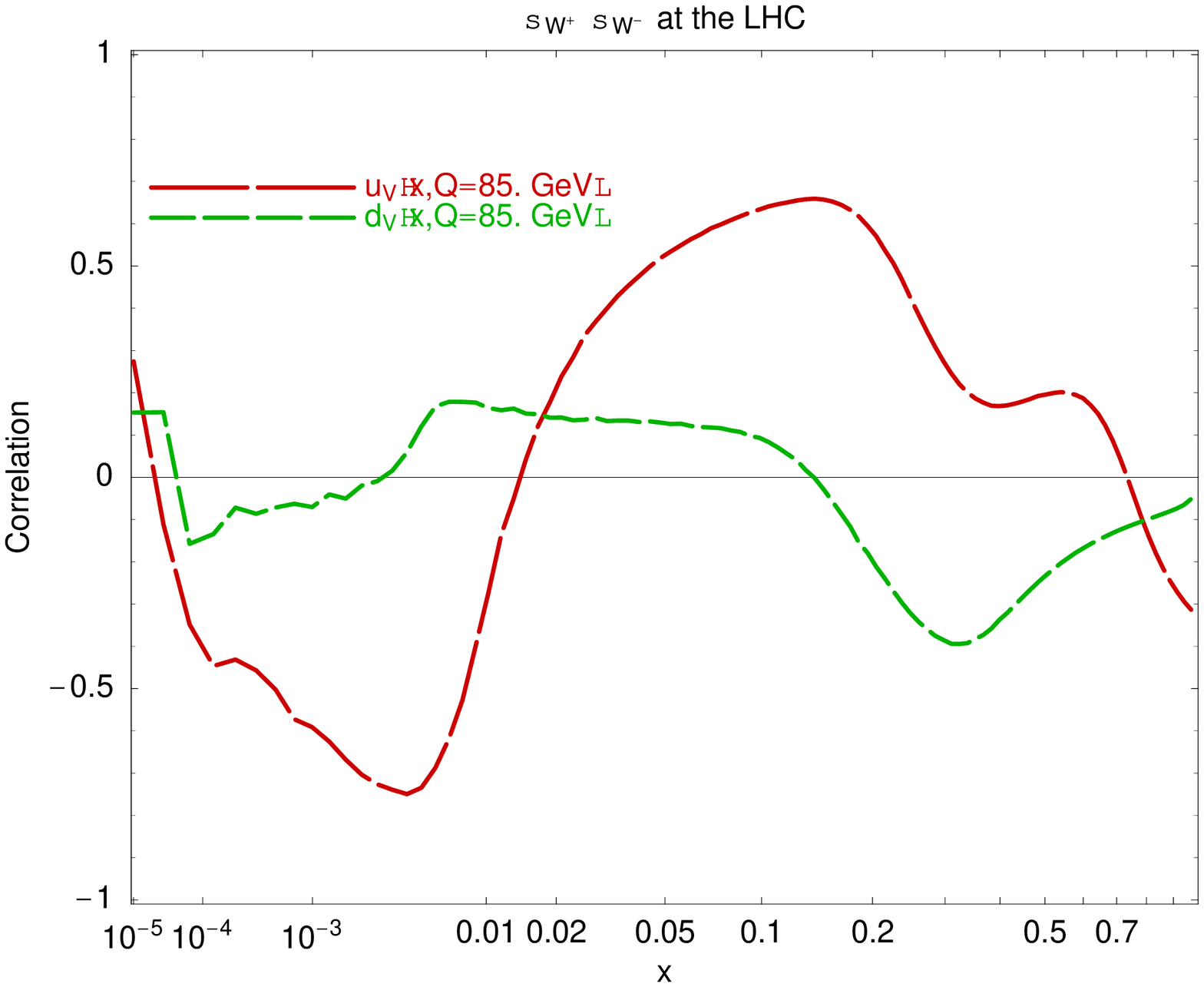}\\
(a)\hspace{2.7in}(b)\par\end{centering}

\caption{(a) CTEQ6.6 $W^{-}$ and $W^{+}$ production cross sections at the
LHC compared with predictions for other PDF sets; (b) correlation
($\cos\varphi$) between the ratio $\sigma_{W^{+}}/\sigma_{W^{-}}$
of $W^{+}$ and $W^{-}$ total cross sections and valence quark PDFs.
\label{fig:WpWmXsecsLHC}}
\end{figure}

The charge dependence of weak boson production at the LHC is explored
by plotting $W^{-}$ vs. $W^{+}$ cross sections in Fig.~\ref{fig:WpWmXsecsLHC}(a).
In this case, the CTEQ6.6 prediction agrees well with the latest MSTW
and AMP sets, while more substantial differences exist with the earlier
MRST and ZEUS PDF sets. Possible intrinsic charm contributions (indicated
by the IC-BHPS and IC-Sea points) would increase both cross sections.

The cross section ratio $\sigma_{W^{-}}/\sigma_{W^{+}}$ is most correlated
with the valence $u$-quark PDF $u_{V}=u-\bar{u}$ at $Q=85$ GeV,
followed by the valence $d$-quark {[}cf.~Fig.~\ref{fig:WpWmXsecsLHC}(b)].
There is a large correlation with $u_{V}(x,Q)$ at $x\approx0.1$
and anticorrelation at $x\approx0.05.$ Other PDF flavors do not demonstrate
pronounced correlation with $\sigma_{W^{-}}/\sigma_{W^{+}}$ and are
not shown in the figure.

\subsection{Top-quark production and gluon uncertainty \label{sub:Top-quark-production}}

The prominent role of gluons in driving the PDF uncertainty in many
processes has been pointed out in the past by noticing that most of
the PDF uncertainty is often generated by a certain PDF eigenvector
sensitive to the gluon parameters. While the abstract eigenvectors
give largely a qualitative insight, the correlation analysis relates
the PDF uncertainty directly to parton distributions for physical
flavors at known $(x,\mu)$. We will now apply the correlation technique to investigate
the uncertainties associated with gluons and heavy quarks in two other
prominent processes, production of top quark-antiquark pairs and single
top quarks.

\subsubsection{Parametrization of $t\bar{t}$ and single-$t$ total cross sections\label{sub:ttbTotalXSecs}}

The inclusive rate for production of top quark-antiquark pairs, $pp\!\!\!{}^{^{(-)}}\rightarrow t\bar{t}X,$
is measured with good precision in the Tevatron Run-2, and an even
more precise measurement is possible with the large event yield expected
at the LHC. Such a measurement can test the PQCD prediction and provide
an alternative method to determine the mass of the top quark \cite{Frixione:1995fj,Cacciari:2003fi}.
At leading order, $t\bar{t}$ pairs are produced via $q\bar{q}$ scattering
(contributing 85\% of the rate at the Tevatron) and $gg$ scattering
(contributing 90\% of the rate at the LHC). At NLO, both the fixed-order
\cite{Nason:1987xz,Nason:1989zy,Beenakker:1988bq,Beenakker:1990maa}
and resummed \cite{Laenen:1991af,Laenen:1993xr,Berger:1995xz,Berger:1996ad,Berger:1997gz,Catani:1996yz,Bonciani:1998vc,Cacciari:2003fi,Kidonakis:1996aq,Kidonakis:1997gm,Kidonakis:2000ui,Kidonakis:2001nj,Kidonakis:2003qe}
cross sections have been computed. In this study, we calculate the
NLO $t\bar{t}$ cross section using CTEQ6.6 PDFs and the MCFM code
\cite{Campbell:2004ch,Campbell:2005bb,Campbell:2000bg}, for three
values of the factorization scale ($\mu=0.5\, m_{t},$ $m_{t},$ and
$2\, m_{t}$). 

Production of single top quarks $pp\!\!\!{}^{^{(-)}}\rightarrow t^{\pm}X$
\cite{Dawson:1984gx,Willenbrock:1986cr,Yuan:1989tc,Cortese:1991fw,Ellis:1992yw,Stelzer:1995mi,Smith:1996ij,Stelzer:1997ns,Mrenna:1997wp,Stelzer:1998ni,Bordes:1992sv,Bordes:1994ki,Ladinsky:1990ut,Moretti:1997ng,Harris:2002md,Sullivan:2004ie,Cao:2004ky,Cao:2004ap,Cao:2005pq,Kidonakis:2006bu,Kidonakis:2007ej}
provides a unique means to measure the $Wtb$ coupling with the goal
of constraining new physics \cite{Tait:2000sh,Tait:1997fe,Yuan:1989tc,Carlson1995,Heinson:1996zm}.
It was recently observed for the first time by the Tevatron D\O\,
\cite{Abazov:2006gd} and CDF Collaborations\ \cite{Wagner:2007nv}.
We will focus on two single-top production channels, proceeding through
$t$-channel and $s$-channel exchanges of charged weak ($W$) bosons.
The $t$-channel $W$ exchange involves bottom-quark scattering $qb\rightarrow q^{\prime}t$
and dominates both the Tevatron and LHC rates. The $s$-channel $W$
exchange is similar to conventional $W$ boson production, but occurs
at larger typical $x$ values (of order $m_{t}/\sqrt{s}$ rather than
$M_{W}/\sqrt{s}$). It may be observable at the Tevatron, but has
a relatively small event rate at the LHC. We compute the NLO cross
sections for single-top production using the programs from Refs.~\cite{Harris:2002md,Sullivan:2004ie}
and \cite{Cao:2004ky,Cao:2004ap,Cao:2005pq}. 

\begin{figure}
\begin{centering}\includegraphics[width=0.5\textwidth,height=286pt,keepaspectratio]{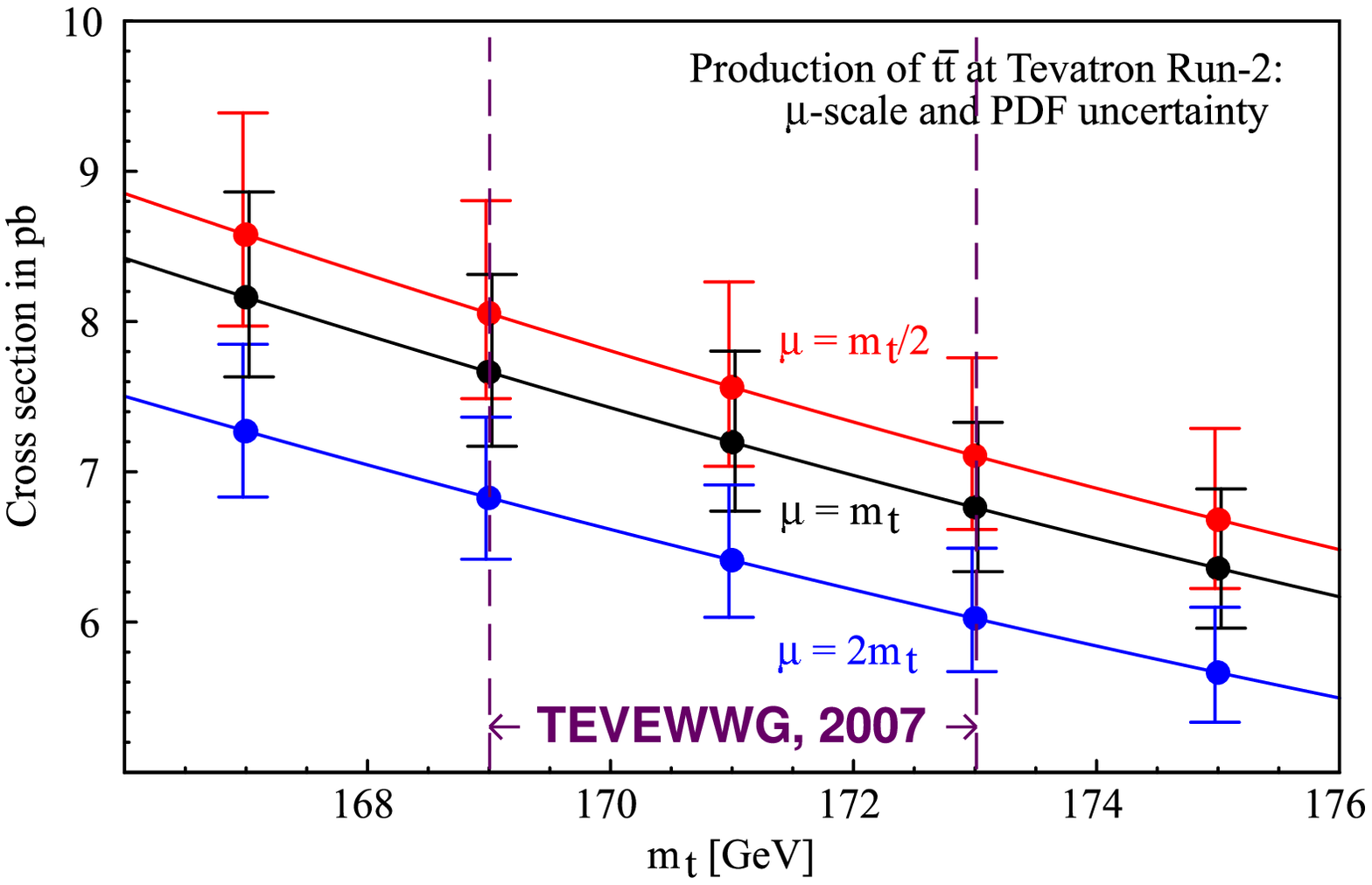}\includegraphics[width=0.5\textwidth,height=286pt,keepaspectratio]{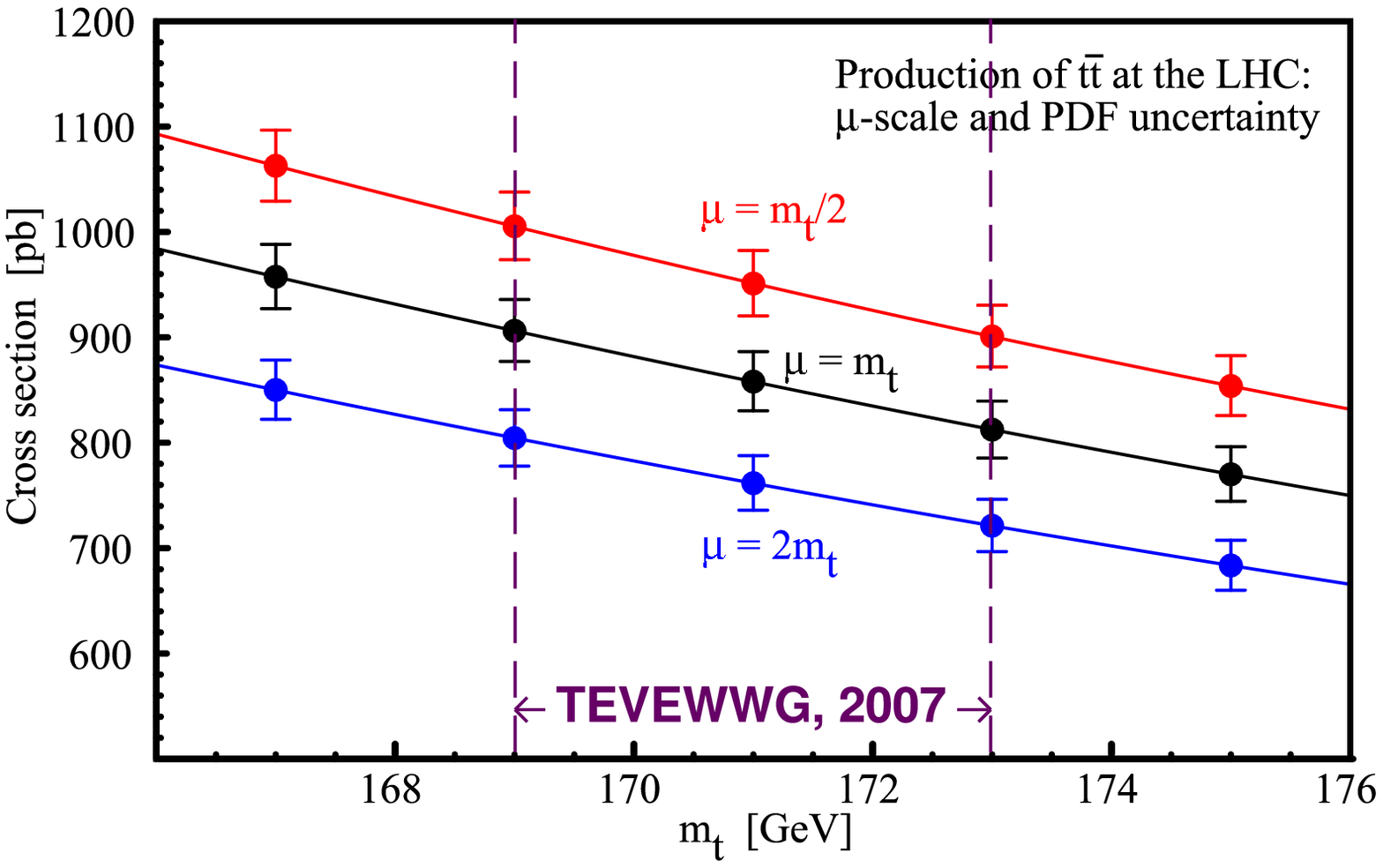}\\
(a)\hspace{2.7in}(b)
\end{centering}
\caption{CTEQ6.6 predictions for inclusive $t\overline{t}$ production at
(a) the Tevatron Run-2 and (b) the LHC, showing the NLO cross section
in pb versus the top-quark mass $m_{t}$. The three curves correspond
to three choices for the factorization scale: $\mu=m_{t}/2,\ $ $m_{t},$
and $2\, m_{t}$. The error bars are the PDF uncertainties. Also shown
is the $1\sigma$ error range from the 2007 world-average experimental
$m_{t}$ value by the Tevatron Electroweak Working Group, $m_{t}=171\pm1.1\pm1.5$
GeV \cite{:2007bxa}. \label{fig:ttbarLHC}}
\end{figure}

For each scale $\mu$, we parametrize the resulting cross sections
in the vicinity of the world-average top mass $m_{t}=171\pm1.1(stat)\pm1.5(syst)$
GeV \cite{:2007bxa} by a function \begin{equation}
\sigma(m_{t},\mu)=A(\mu)+B(\mu)\left(m_{t}-171\right)+C(\mu)\left(m_{t}-171\right)^{2},\label{sigmatop}\end{equation}
where the units of $\sigma$ and $m_{t}$ are picobarn and GeV. The
variation with respect to the reference cross section (corresponding
to $\mu=m_{t})$ gives the NLO scale dependence $\Delta_{\mu}(m_{t})$,
discussed in detail in Section~\ref{sub:ttbLuminosityHiggs}. The
uncertainty in $\sigma$ due to the variation of $m_{t}$ within the
experimentally allowed range gives the mass dependence, denoted by
$\Delta_{m}$. We also calculate the relative PDF uncertainties, denoted
by $\Delta_{PDF}(m_{t})$. The values of $\Delta_{\mu}$, $\Delta_{PDF},$
and coefficients $A,$ $B,$ $C$ are listed in Table~\ref{tab:ttbarABC}.
In single-top cross sections, we set $C=0.$ 
A plot of this parametric dependence in $t\bar{t}$ production at
the Tevatron and LHC is shown in Fig.~\ref{fig:ttbarLHC}.
The correlation cosines
between $t\bar{t},$ single-top, $W,$ and $Z$ cross sections are
listed in Tables~\ref{tab:CorrsStdCandles} and \ref{tab:CorrsSingleTop}.

\subsubsection{PDF-induced correlations \label{sub:ttbCorr}}

\begin{figure}[p]
\begin{centering}\includegraphics[width=0.5\textwidth,height=0.265\textheight,keepaspectratio]{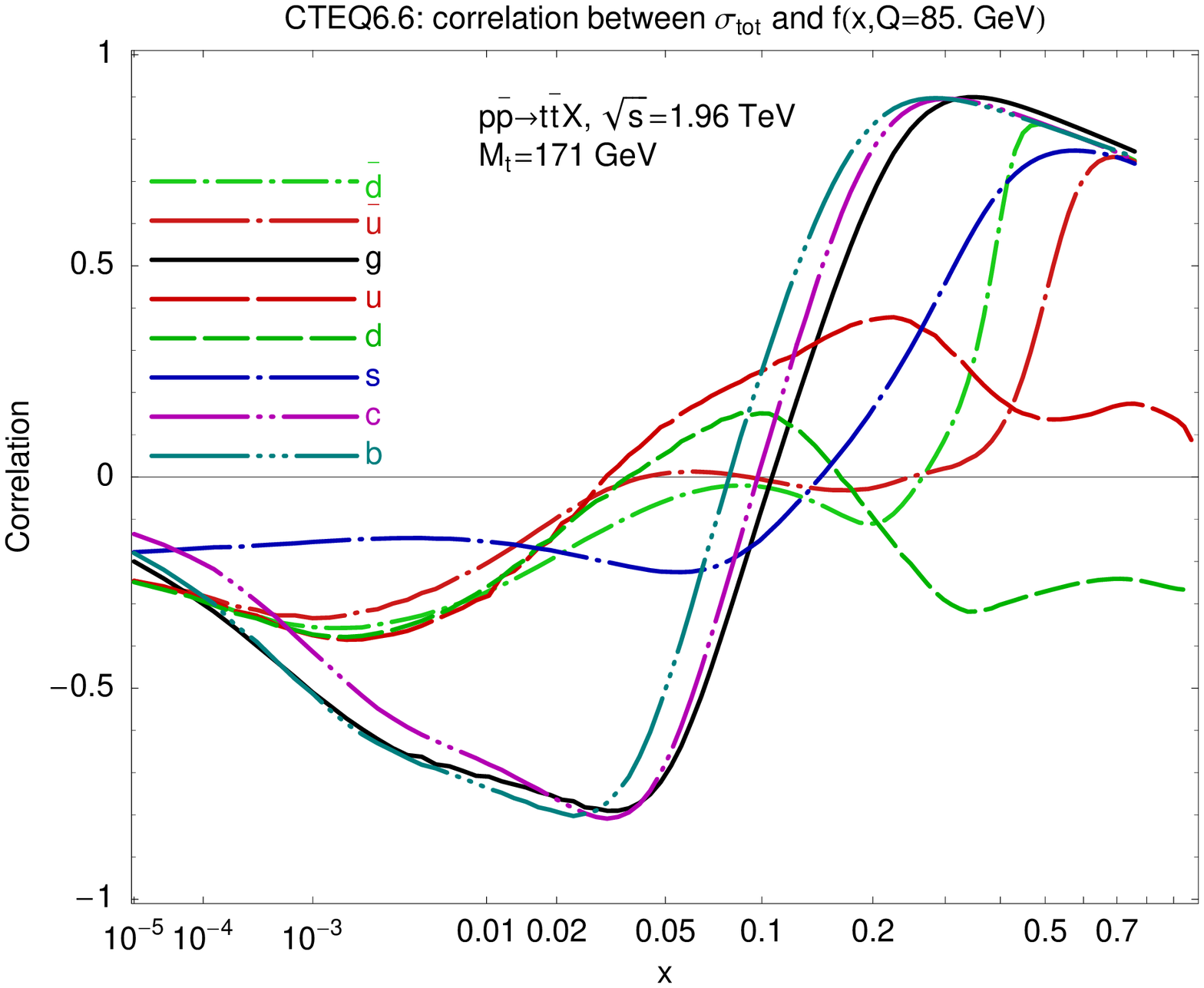}\includegraphics[width=0.5\textwidth,height=0.265\textheight,keepaspectratio]{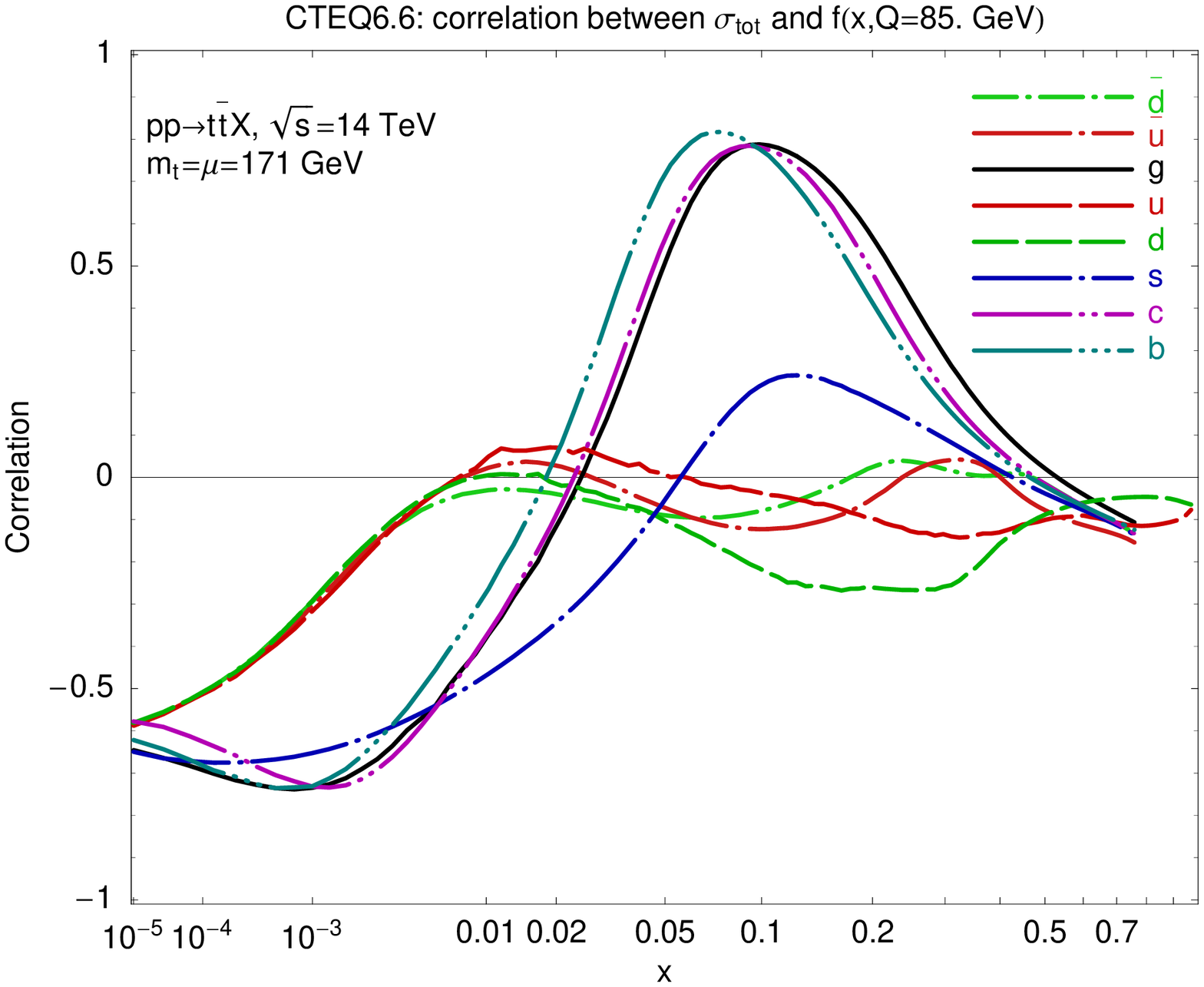}\\
(a)\hspace{2.7in}(b)\par\end{centering}

\begin{centering}\includegraphics[width=0.5\textwidth,height=0.265\textheight,keepaspectratio]{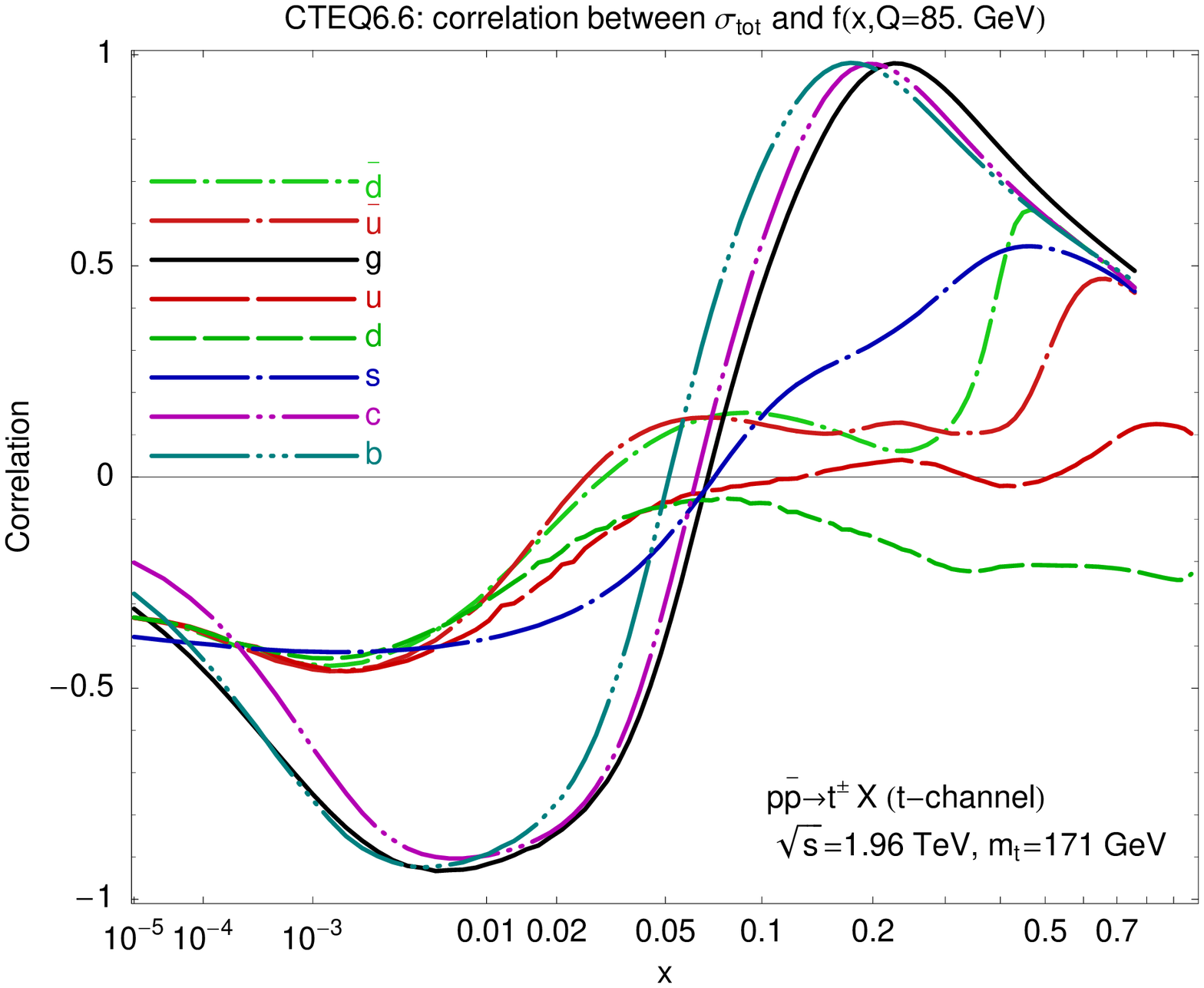}\includegraphics[width=0.5\textwidth,height=0.265\textheight,keepaspectratio]{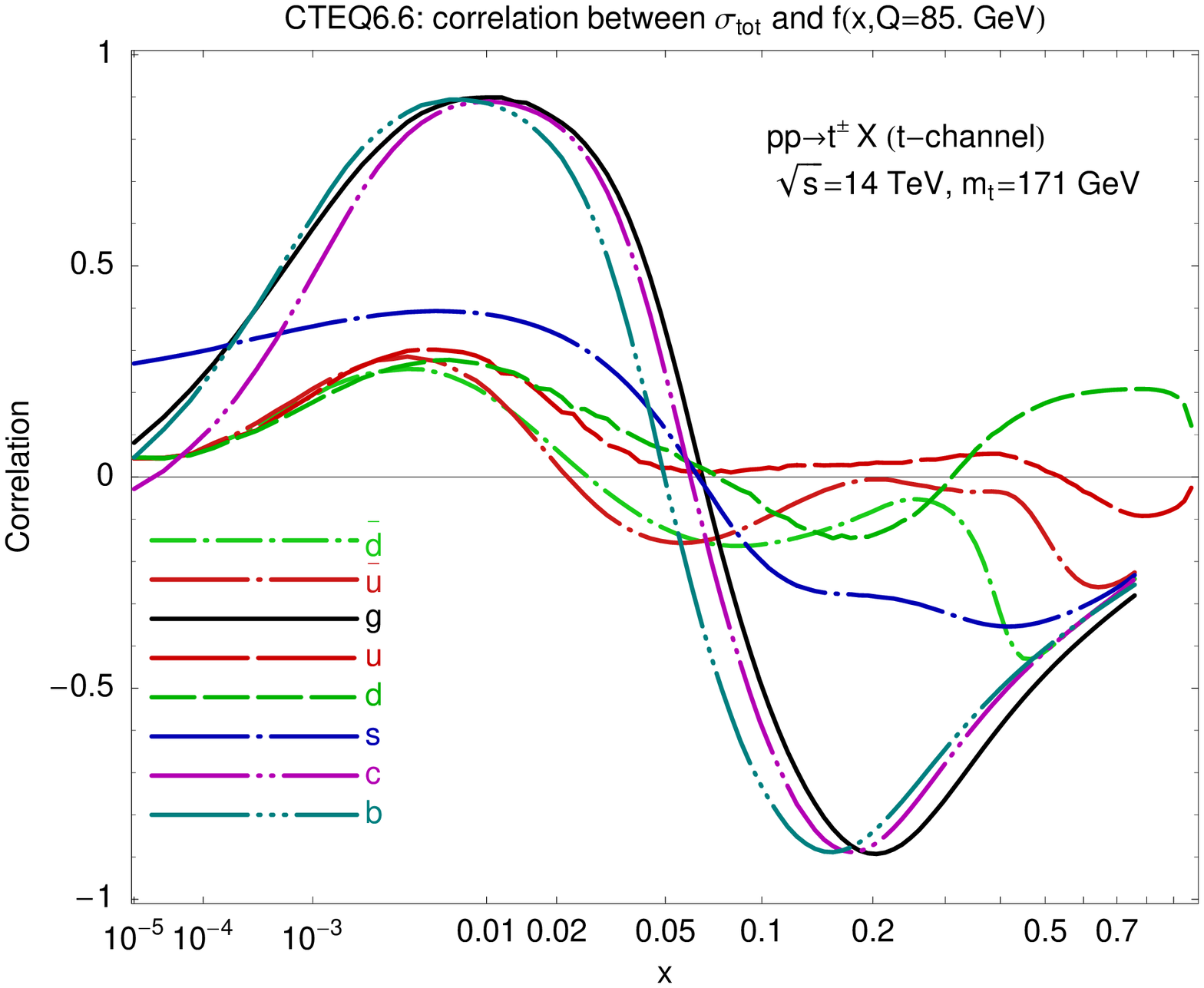}\\
(c)\hspace{2.7in}(d)\par\end{centering}

\begin{centering}\includegraphics[width=0.5\textwidth,height=0.265\textheight,keepaspectratio]{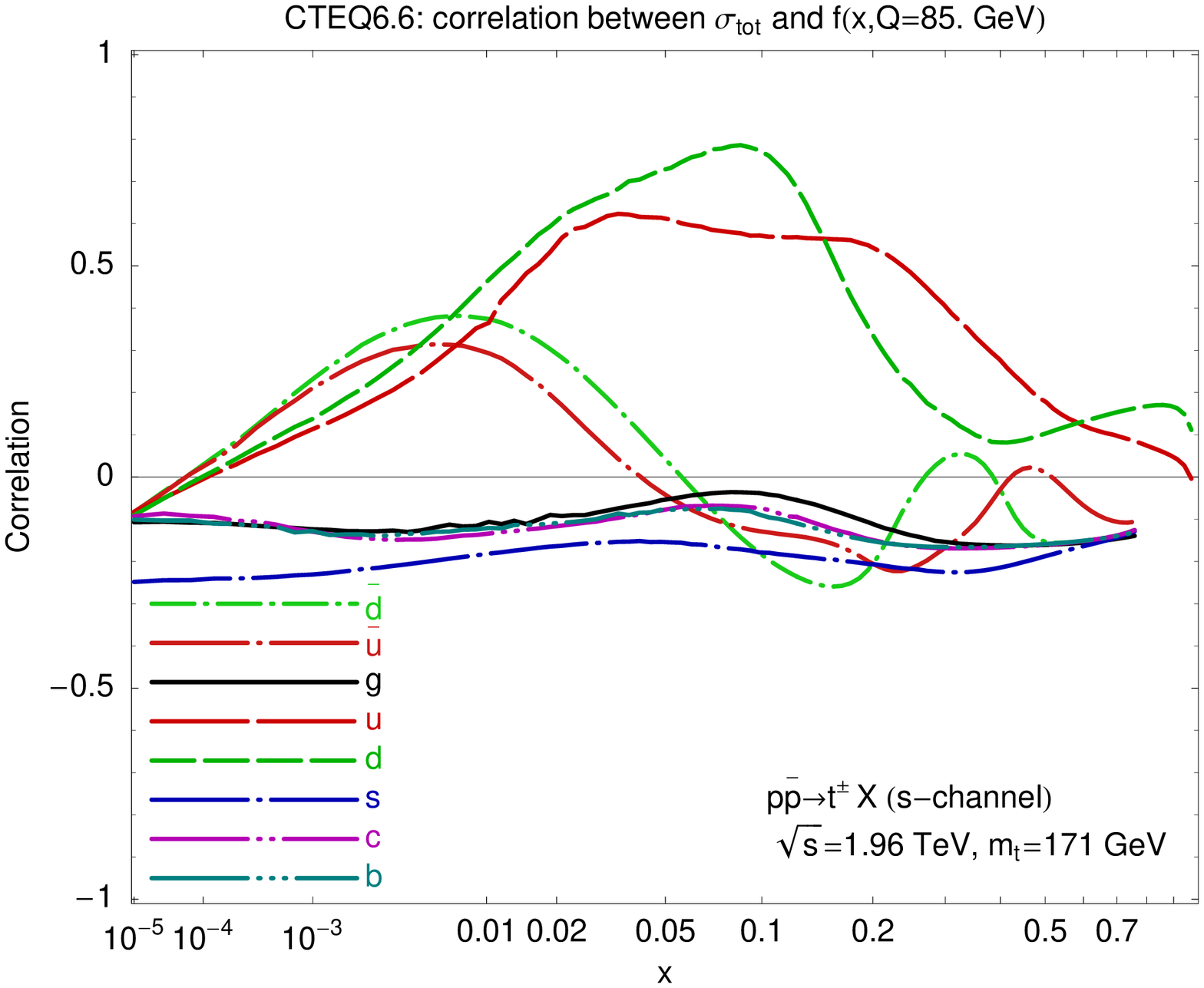}\includegraphics[width=0.5\textwidth,height=0.265\textheight,keepaspectratio]{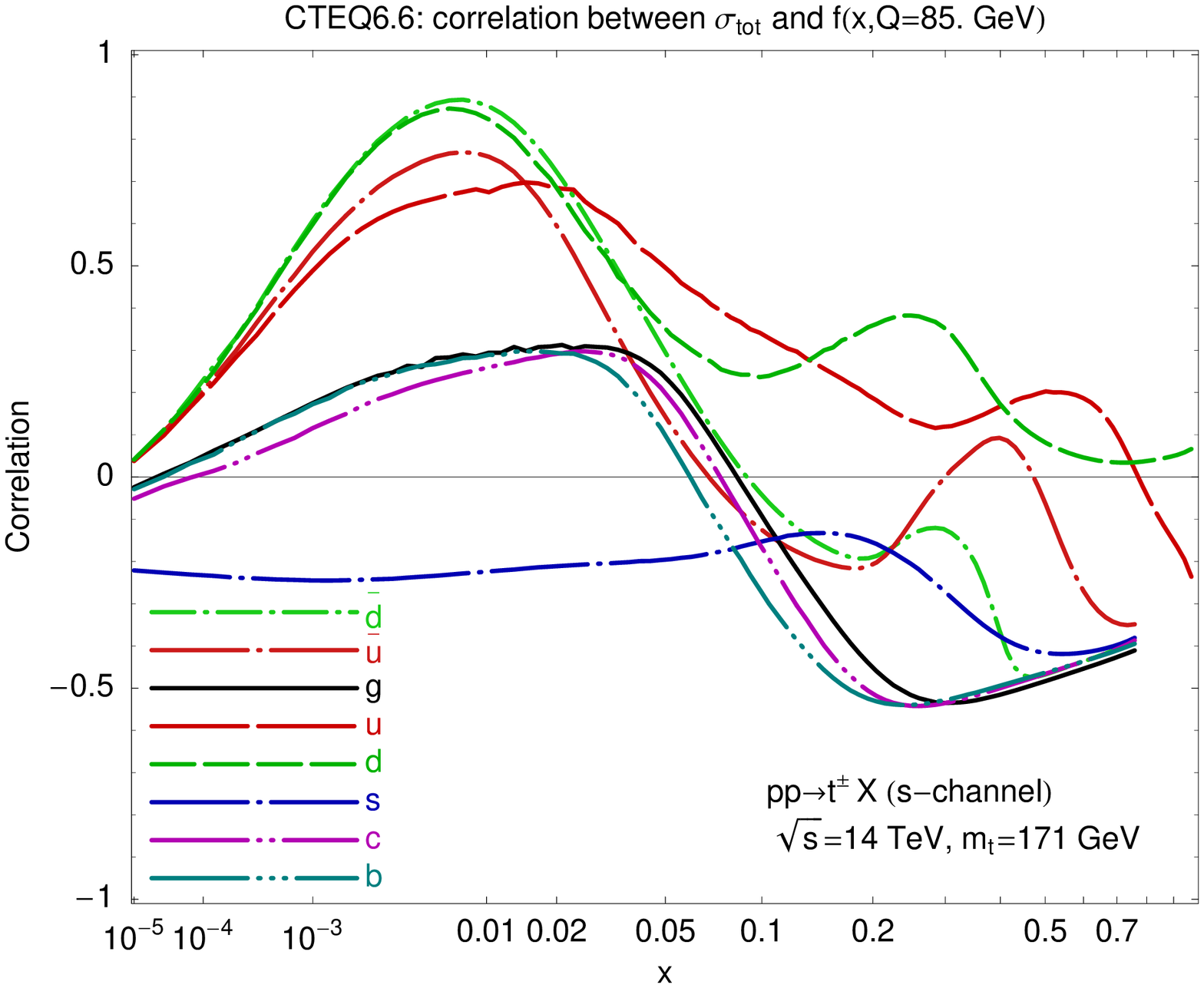}\\
(e)\hspace{2.7in}(f)\par\end{centering}

\caption{PDF-induced correlations ($\cos\varphi$) between the total cross
sections for $t\bar{t}$ and single-top production, and PDFs of various
flavors, plotted as a function of $x$ for $Q=85$ GeV: Tevatron Run-2
(left column); LHC (right column). \label{fig:ttb_sngl_top_corr}}
\end{figure}

The PDF dependence of the top-quark cross sections follows a few non-trivial
trends, which can be understood by studying $x$-dependent correlations
between the top-quark cross sections and PDFs presented in Fig.~\ref{fig:ttb_sngl_top_corr}.
Our discussion will also refer to Tables~\ref{tab:CorrsStdCandles}-\ref{tab:CorrsSingleTop}.

\begin{figure}
\begin{centering}\includegraphics[width=0.5\textwidth]{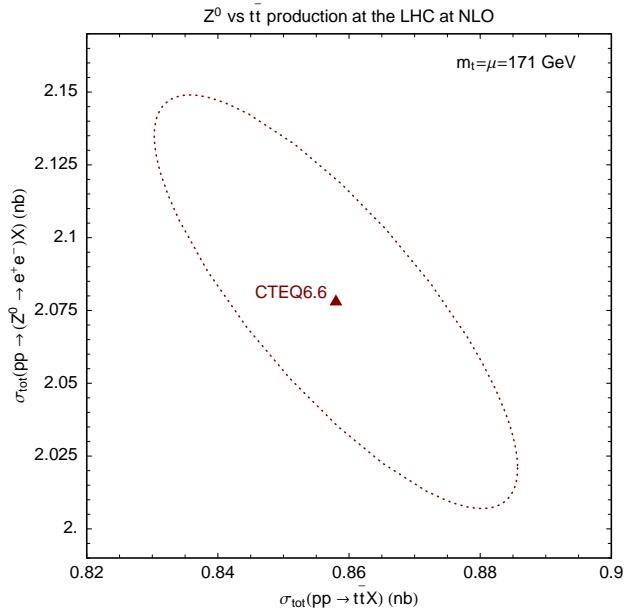}\par\end{centering}

\caption{Correlation ellipse for $t\bar{t}$ and $Z$ boson NLO total cross sections
at the LHC. \label{fig:zttb_corrLHC}}
\end{figure}

\begin{enumerate}
\item Contrary to the naive expectation, the main PDF uncertainty in $t\bar{t}$
production at the Tevatron is not associated with the leading $q\bar{q}$
scattering channel. Rather, the uncertainty is mostly correlated with
the gluon PDF probed at $x\approx0.3$, as clearly shown by the $t\bar{t}$-PDF
correlations in Fig.~\ref{fig:ttb_sngl_top_corr}(a). At such $x$
values, the quark PDFs are tightly constrained, resulting in a small
uncertainty in the leading $q\bar{q}$ channel, while the gluons are
poorly known, resulting in a very large uncertainty in the subleading
$gg$ channel. The net result is a substantial total PDF uncertainty,
$\Delta_{PDF}=7.4\%$, in the Tevatron $t\bar{t}$ cross section,
contributed mostly by $gg$ scattering. By the momentum sum rule,
the Tevatron $t\bar{t}$ cross section is strongly anti-correlated
with gluon scattering at $x\sim0.05.$ It does not exhibit a strong
(anti-)correlation with $W$ or $Z$ production at either collider,
cf.~Table~\ref{tab:CorrsStdCandles}. 
\item The Tevatron $t$-channel single-top cross section is mostly correlated
with the $b$-quark PDF at $x\sim0.2$, as illustrated by Fig.~\ref{fig:ttb_sngl_top_corr}(c).
It has a substantial PDF uncertainty, $\Delta_{PDF}=$10.3\%. It is
correlated with the Tevatron $t\bar{t}$ cross section ($\cos\varphi=0.81$)
and anticorrelated with $Z,$ $W$ production at the LHC ($\cos\varphi=-0.82$
and $-0.79$) through the shared correlation with the gluon at large
$x$. 
\item At the LHC {[}Figs.~\ref{fig:ttb_sngl_top_corr}(b) and (d)], the
$t\bar{t}$ and $t$-channel single-top cross sections are also mostly
correlated with $g,$ $c,$ and $b$ PDFs, which, however, are well-constrained
at $x\sim0.05$ and $0.01$, typical values in this case. The PDF
uncertainty is of order 3\% in both processes. The LHC $t\bar{t}$
cross sections are anticorrelated with the $Z$ and $W$ cross sections.
The $t\bar{t}-Z$ and $t\bar{t}-W$ correlation cosines are large
and negative: $\cos\varphi=-0.8$ and $-0.74$, respectively. The
strong anticorrelation reveals itself in the shape of the $t\bar{t}-Z$
ellipse, plotted in Fig.~\ref{fig:zttb_corrLHC}(a) by using the
PDF error parameters in Table~\ref{tab:CorrsStdCandles}. A similar
anticorrelation exists between the $t\bar{t}$ and $W$ cross sections
\cite{Campbell:2006wx}. Interestingly enough, the LHC $t-$channel
single-top cross section is only mildly anti-correlated with $t\bar{t}$
production and mildly correlated with $W$ and $Z$ production {[}see
Table~\ref{tab:CorrsSingleTop}], in contrast to the Tevatron. 
\item At both colliders, the $s$-channel PDF uncertainty is of order 3\%
and correlated mostly with $u\!\!\!{}^{^{(-)}}$ and $d\!\!\!{}^{^{(-)}}$
PDFs {[}cf.~Figs.~\ref{fig:ttb_sngl_top_corr}(e) and (f)], as well
as with the Tevatron $Z,$ $W$ rates. Remarkably, the LHC $s$-channel
single-top cross section is not particularly correlated with the LHC
$W,$ $Z$ cross sections, despite its similarities with $W$ boson
production. This peculiarity is due to large $x$ values $(x\sim m_{t}/\sqrt{s}\sim0.01)$
typical for the $s$-channel single-top production. At such $x,$
charm and bottom initial-state contributions are relatively small
and do not affect single-top production as much as $W$ boson production,
hence preventing the gluon-driven PDF uncertainty from contributing
sizably to the $s$-channel cross section. 
\item The improved evaluation of heavy-quark terms in the CTEQ6.6 PDFs reduces
the Tevatron (LHC) $t\bar{t}$ cross sections by 4\% (3.5\%) compared
to CTEQ6.1. The CTEQ6.6 Tevatron $t$-channel single-top cross section
is about 6\% smaller than the CTEQ6.1 cross section. The other three
CTEQ6.6 single-top cross sections differ from the CTEQ6.1 cross sections
by less than 2\%. 
\end{enumerate}

\subsection{$t\bar{t}$ production as a standard candle; Higgs boson production
\label{sub:ttbLuminosityHiggs}}

To recap the previous section, at the Tevatron $t\bar{t}$ production
is strongly correlated with single-top production. At the LHC, it
is strongly anticorrelated with $Z,$ $W$ production. Precise measurements
of $t\bar{t}$ rates could provide valuable constraints on the gluon
and heavy-quark PDFs, if the associated theoretical and experimental
uncertainties are each reduced below $3-5\%$. These measurements
would help bring down the PDF error in many processes and supply an
alternative way to monitor the collider luminosity. If a cross section
is anti-correlated with $W$ and $Z$ production, it could be normalized
to the $t\bar{t}$ cross section. The PDF error will be suppressed
in such a cross section ratio, in contrast to the ratio with the $W$
or $Z$ cross section. Other systematic errors may cancel better,
too, if the process shares common elements with $t\bar{t}$ production.

Let us consider some specific examples. The PDF uncertainties for
$Z,$ $W^{\pm},$ $t\bar{t},$ and Higgs boson ($h^{0}$) production
via gluon fusion at the LHC, with Higgs mass $m_{h}=500$ GeV, are
3.4\%, 3.5\%, 3.2\%, and 4\%, respectively. The correlation cosines
are $0.956$ for $Z$ and $W^{\pm}$ cross sections, $0.98$ for $h^{0}$
and $t\bar{t}$ cross sections, and $-0.87$ for $h^{0}$ and $Z$
cross sections. By Eq.~(\ref{dfrat}) the PDF uncertainty on the
$Z/W^{\pm},$ $h^{0}/t\bar{t},$ $h^{0}/Z$ cross section ratios are
1.3\%, 1.5\%, and 7.2\%, i.e., the correlated cross sections produce
ratios with the smallest PDF uncertainties.

The viability of precise $t\bar{t}$ production measurements can be
examined by studying theoretical uncertainties on the $t\bar{t}$
cross section presented in Table~\ref{tab:ttbarABC}. At present,
the scale dependence $\Delta_{\mu}$ is larger than the PDF uncertainty
$\Delta_{PDF}$ and top-mass uncertainty $\Delta_{m}$ at both colliders,
suggesting that higher-order (NNLO) corrections have a tangible impact
on the $t\bar{t}$ rate.%
\footnote{The scale dependence of the NLO $t\bar{t}$ cross section may be reduced
by threshold resummation \cite{Catani:1996yz,Bonciani:1998vc,Cacciari:2003fi,Kidonakis:2003qe},
which includes higher-order logarithmic terms that enforce renormalization
group invariance. In our study the scale dependence is viewed as an
estimate of all NNLO contributions, including potentially sizable
higher-order terms not associated with the threshold logarithms. A
more conservative estimate of full NNLO effects is provided by the
scale dependence of the fixed-order NLO cross section~(\ref{sigmatop}),
rather than that of the threshold-resummed NLO cross section.%
} The NNLO contributions will be computed in the near future \cite{Czakon:2007wk},
which will likely reduce the associated uncertainty to a few percent. 

The top-mass uncertainty $\Delta_{m}$ can be brought down to 2-3\%
by measuring $m_{t}$ with accuracy of order 1 GeV, as planned by
the Tevatron experiments. Further advancements can possibly improve
accuracy in the treatment of heavy-quark mass effects at NNLO to about
a percent level. The current (not related to the luminosity) experimental
systematic error for the $t\bar{t}$ cross section at the Tevatron
is 8\% \cite{CDFNote8148}, with further improvements likely. A similar
systematic error of order 5\% may be feasible at the LHC. 

\begin{figure}
\includegraphics[width=0.8\columnwidth]{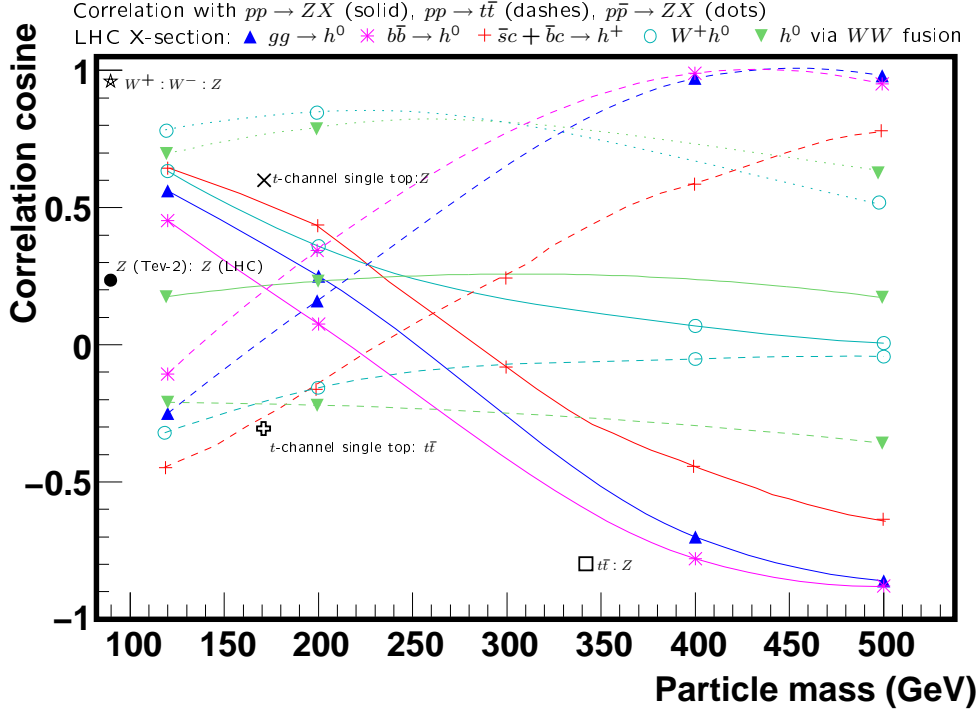}

\caption{The correlation cosine $\cos\varphi$ for Higgs boson searches at
the LHC with respect to $Z$ boson production at the LHC (solid) and
Tevatron (dots), and $t\bar{t}$ production at the LHC (dashes), plotted
as a function of Higgs mass. Separate markers denote correlations
of $W$, $t$-channel single-top cross sections at the LHC and $Z$ cross section 
at the Tevatron with respect to $Z$ and $t\bar t$ cross sections at
the LHC. \label{fig:Correlation-cosine-vs-mass}}
\end{figure}

The complementarity of constraints on Higgs boson searches from $t\bar{t}$
production at the LHC and $Z$ boson production at the Tevatron and
LHC is illustrated by Fig.~\ref{fig:Correlation-cosine-vs-mass},
showing the correlation cosines ($\cos\varphi$) between Higgs, $Z,$
and $t\bar{t}$ production cross sections as a function of Higgs mass.
These cosines are shown by lines, in addition to separate markers
corresponding to correlations between $W,$ $Z,$ and top production
processes discussed above. The relevant cross sections are collected
in Table~\ref{tab:CorrsStdCandles} for $W,$ $Z,$ and $t\bar{t}$
production, and in Table~\ref{tab:HiggsXSecsHC} for Higgs production
processes. 

As discussed earlier, there is a very large correlation between $Z$
boson, $W^{+}$ boson, and $W^{-}$ boson production at the LHC, and
a strong anticorrelation between $Z$ boson and $t\bar{t}$ production.
Only a mild correlation exists between $Z$ production at the Tevatron
and $Z$ production at the LHC. There is a moderate correlation between
$Z$ production and the production of a light (120 GeV) Higgs boson
through $gg$ fusion, but this becomes a strong anticorrelation as
the mass of the Higgs boson increases, as the gluons are in a similar
$x$ range as those responsible for $t\overline{t}$ production. Associated
Higgs boson production ($Wh^{0}$) is strongly correlated with $Z$
production for low Higgs masses but becomes decorrelated for higher
masses.%
\footnote{Similarly, $Wh^{0}$ associated production ($m_{h}<200$ GeV) and
$Z$ boson production are strongly correlated at the Tevatron. %
} There is only a mild correlation between the production of a Higgs
boson through vector boson fusion and $Z$ production over a wide
range of $m_{h}$.

The correlation curves with respect to $t\overline{t}$ production
basically form a mirror image to the previous curves, since $t\overline{t}$
pairs at the LHC are predominantly produced via $gg$ fusion at large
$x$. Thus, for example, the production of a Higgs boson through $gg$
fusion goes from a mild anti-correlation with $t\bar{t}$ for low
Higgs masses to a high correlation for large Higgs masses.

PDF-induced correlations may follow a different pattern in other two-Higgs
doublet models (2HDM). For example, the MSSM Higgs boson production
$c\bar{s}+c\bar{b}\rightarrow h^{+}$ is not particularly correlated
with $t\bar{t}$ production at $m_{h}\gtrsim500$ GeV because of the
uncorrelated contribution from the anti-strangeness PDF (cf.~Fig.~\ref{fig:Correlation-cosine-vs-mass}).
The $c\bar{s}\rightarrow h^{+}$ channel is absent in other 2HDM,
such as the effective weak-scale 2HDM induced by top-color dynamics
\cite{Balazs:1998sb}, where the scattering proceeds entirely via
$c\bar{b}\rightarrow h^{+}$ at the Born level. In those models the
correlation of the $t\bar{t}$ cross section with the $h^{+}$ cross
section is very strong at large $m_{h}$ ($\cos\varphi=0.98$ for
$m_{h}=500$ GeV).

Higgs boson production at the LHC is affected by the new features of CTEQ6.6
PDFs in several ways. Table~\ref{tab:Higgsproduction} lists the
relative difference $\Delta_{GM}\equiv\sigma_{6.1}/\sigma_{6.6}-1$
between the CTEQ6.1 and CTEQ6.6 cross sections, as well as CTEQ6.6
PDF uncertainties $\Delta_{PDF}$ for Higgs boson production in a
mass range $m_{h}=100-800$ GeV. In processes dominated by light-quark
scattering (vector boson fusion, $W^{\pm}h$, $Z^{0}h$, and $Ah^{\pm}$),
the most tangible differences between CTEQ6.6 and CTEQ6.1 (compared
to the PDF uncertainty) occur at $m_{h}$ of order 100 GeV, where
$\Delta_{GM}$ is close in magnitude to $\Delta_{PDF}$, reflecting
the enhancement in CTEQ6.6 $u$ and $d$ PDFs at $x=10^{-3}-10^{-2}$.
In gluon-gluon fusion $gg\rightarrow h$, the difference between CTEQ6.6
and 6.1 is well within the PDF uncertainty ($\Delta_{GM}<\Delta_{PDF}$),
although $\Delta_{GM}$ becomes comparable to $\Delta_{PDF}$ for
heavy Higgs masses ($m_{h}\gtrsim500$ GeV). Similarly, $\Delta_{GM}$
is smaller than $\Delta_{PDF}$ in heavy-quark scattering $c\bar{b}\rightarrow h^{+}.$
The most striking differences occur in $c\bar{s}\rightarrow h^{+},$
because the CTEQ6.1 and CTEQ6.6 strangeness distributions disagree
by a large amount in most of the $x$ range.

\section{Conclusion\label{sec:Conclusion}}

The new CTEQ6.6 global analysis incorporates the latest improvements
in the perturbative QCD treatment of $s,$ $c,$ and $b$ quark PDFs
introduced in Refs.~\cite{Tung:2006tb,Lai:2007dq,Pumplin:2007wg}.
It predicts substantial modifications (comparable to, or exceeding in magnitude the
NNLO contributions) in high-energy electroweak precision cross sections.
Theoretical improvements of this kind must be accompanied by the development
of efficient strategies to understand and reduce the remaining uncertainties
in the PDF parameters. This work presents a novel correlation analysis,
a technique based on the Hessian method that links the PDF uncertainty
in a hadronic cross section to PDFs for \emph{physical} parton flavors
at well-defined $(x,\mu)$ values. 

We apply the correlation analysis to reveal and explain several regularities
observed in the PDF dependence, such as strong sensitivity of $W,$
$Z$ production at the LHC and $t\bar{t}$ production at the Tevatron
(processes dominated by $q\bar{q}$ scattering) to uncertainties in
the gluon and heavy-quark PDFs; the leading role played by the strangeness
distribution $s(x,\mu)$ in the PDF uncertainty of the ratio $r_{ZW}=\sigma_{Z}/\sigma_{W}$
of the LHC $Z$ and $W$ cross sections; and intriguing PDF-induced
anticorrelations between the LHC $Z,$ $W$ cross sections and processes
dominated by large-$x$ gluon-scattering, such as heavy Higgs boson
production via gluon fusion. 

In our study we identify pairs of hadronic cross sections with strongly
correlated or anticorrelated PDF dependence, i.e. with the correlation
cosine $\cos\varphi$ close to 1 or -1. Such pairs are especially
helpful for constraining the PDF uncertainties, in view that a combination
of the existing CTEQ6.6 and upcoming LHC constraints on one cross
section in the pair is guaranteed to reduce substantially the PDF
uncertainty on the second cross section {[}cf.~the discussion accompanying
Eq.~(\ref{PCTEQLHC})]. In addition, ratios of correlated (but not
anticorrelated) cross sections have greatly reduced PDF uncertainty,
as follows from Eq.~(\ref{dfrat}). For this reason, it is beneficial
to normalize an LHC cross section to a standard candle cross section
with which it has a large PDF-induced correlation.

We point out a potentially valuable role of precise measurements of
$t\bar{t}$ cross sections at the Tevatron and LHC for constraining
the gluon PDF at large $x$ and normalizing the LHC cross sections
that are anticorrelated with $W$ and $Z$ boson production. If both
theoretical and experimental uncertainties on $t\bar{t}$ cross sections
are reduced to a level of 3-5\%, as may become possible in the near
future, the $t\bar{t}$ cross sections would provide an additional
standard candle observable with useful complementarity to $Z$ and
$W$ boson cross sections. These measurements will be essential for
reducing theoretical uncertainties in single-top and Higgs boson production,
and for constructing cross section ratios with small PDF uncertainties. 

\noindent \textbf{\vspace*{\smallskipamount}
~}\\
\textbf{Acknowledgments}

\noindent
We thank J.~Smith for providing a computer program to verify
the $t\bar{t}$ production cross sections; A.~Cooper-Sarkar, N.~Kidonakis,
M.~Mangano, J.~Stirling, and CTEQ members for helpful
communications; and E.~L.~Berger for the critical reading 
of the manuscript.  We are especially grateful 
to R. Thorne for a valuable comment 
about the strangeness distribution at small $x$ that resulted 
in improvements in the revised paper, and for other discussions.

This work was supported in part by the U.S. National
Science Foundation under awards PHY-0354838, PHY-0555545, and PHY-0551164;
by the U.S. Department of Energy under Grant No. DE-FG03-94ER40837; and by
the National Science Council of Taiwan under grant NSC-95-2112-M-133-001.
The work of P.~M.~N. on an early version of the manuscript at the Argonne National
Laboratory  was supported by the U.S. Department of Energy, 
Division of High Energy Physics, under Contract DE-AC02-06CH11357. 
P.~M.~N. acknowledges the hospitality of Kavli Institute 
for Theoretical Physics at the University of California in Santa Barbara 
during the final stage of this work. 

\newpage
\begin{table}[H]

\caption{Total cross sections $\sigma$, PDF-induced errors $\Delta\sigma$,
and correlation cosines $\cos\varphi$ for $Z^{0}$, $W^{\pm}$, and
$t\bar{t}$ production at the Tevatron Run-2 (Tev2) and LHC, computed
with CTEQ6.6 PDFs.\label{tab:CorrsStdCandles}}

\begin{tabular}{|c|c|c|c|c|c|c|}
\hline 
$\sqrt{s}$&
Scattering&
$\sigma,\Delta\sigma$&
\multicolumn{4}{c|}{Correlation  $\cos\varphi$ with}\tabularnewline
(TeV)&
process&
(pb)&
$Z^{0}$ (Tev2)&
$W^{\pm}$(Tev2)&
$Z^{0}$ (LHC)&
$W^{\pm}$ (LHC)\tabularnewline
\hline
\hline 
&
$p\bar{p}\rightarrow(Z^{0}\rightarrow\ell^{+}\ell^{-})X$&
241(8)&
1&
\textbf{0.987}&
0.23&
0.33\tabularnewline
1.96&
$p\bar{p}\rightarrow{(W}^{\pm}\rightarrow\ell\nu_{\ell})X$&
2560(40)&
\textbf{0.987}&
1&
0.27&
0.37\tabularnewline
&
$p\bar{p}\rightarrow t\bar{t}X$&
7.2(5)&
-0.03&
-0.09&
-0.52&
-0.52\tabularnewline
\hline 
&
$pp\rightarrow{(Z}^{0}\rightarrow\ell^{+}\ell^{-})X$&
2080(70)&
0.23&
0.27&
1&
\textbf{0.956}\tabularnewline
&
$pp\rightarrow{(W}^{\pm}\rightarrow\ell\nu)X$&
20880(740)&
0.33&
0.37&
\textbf{0.956}&
1\tabularnewline
14&
$pp\rightarrow(W^{+}\rightarrow\ell^{+}\nu_{\ell})X$&
12070(410)&
0.32&
0.36&
\textbf{0.928}&
\textbf{0.988}\tabularnewline
&
$pp\rightarrow(W^{-}\rightarrow\ell^{-}\bar{\nu}_{\ell})X$&
8810(330)&
0.33&
0.38&
\textbf{0.960}&
\textbf{0.981}\tabularnewline
&
$pp\rightarrow t\bar{t}X$&
860(30)&
-0.14&
-0.13&
\textbf{-0.80}&
\textbf{-0.74}\tabularnewline
\hline
\end{tabular}
\end{table}

\begin{table}[H]

\caption{The fitting parameters $(A,B,C)$ for the parametric form (\ref{sigmatop})
of the CTEQ6.6M total cross section for inclusive $t\bar{t}$ and
single-top production at the Tevatron and LHC, evaluated at NLO in
the QCD coupling strength. Also shown are the relative scale and PDF
errors, $\Delta_{\mu}$ and $\Delta_{PDF}$ at $m_{t}=171$ GeV.  \label{tab:ttbarABC}}

\begin{tabular}{|c|c||>{\centering}p{0.6in}|>{\centering}p{0.6in}|>{\centering}m{0.6in}||>{\centering}p{0.6in}|>{\centering}m{0.6in}|>{\centering}m{0.6in}|}
\hline 
\multicolumn{2}{|c|}{}&
\multicolumn{3}{c||}{$p\bar{p}\rightarrow TX$ ($\sqrt{s}=1.96$ TeV)}&
\multicolumn{3}{c|}{$pp\rightarrow TX$ ($\sqrt{s}=14$ TeV)}\tabularnewline
\hline 
Final state&
Parameter&
$\mu=m_{t}/2$ &
$\mu=m_{t}$ &
$\mu=2\, m_{t}$ &
$\mu=m_{t}/2$ &
$\mu=m_{t}$ &
$\mu=2\, m_{t}$ \tabularnewline
\hline 
&
$A$ {[}pb] &
7.546 &
7.197 &
6.412 &
951.2 &
857.9 &
761.6 \tabularnewline
&
$B$ {[}pb$\cdot$GeV$^{-1}$] &
-0.237 &
-0.225 &
-0.201 &
-26.12 &
-23.43 &
-20.81 \tabularnewline
\multicolumn{1}{|c|}{$T=t\bar{t}$}&
$C$ {[}pb$\cdot$GeV$^{-2}$] &
0.0041 &
0.0039 &
0.0034 &
0.44 &
0.37 &
0.33 \tabularnewline
\cline{2-2} \cline{3-3} \cline{4-4} \cline{5-5} \cline{6-6} \cline{7-7} \cline{8-8} 
\multicolumn{1}{|c|}{}&
$\Delta_{\mu}(m_{t}=171)$ &
$+5\%$ &
reference &
$-11\%$ &
$+11\%$ &
reference &
$-11\%$ \tabularnewline
\cline{2-2} \cline{3-5} \cline{6-8} 
\multicolumn{1}{|c|}{}&
$\Delta_{PDF}(m_{t}=171)$&
\multicolumn{3}{c|}{\textbf{$_{-6.4}^{+8.4}$ (7.4)\%}}&
\multicolumn{3}{|c|}{$_{-3.2}^{+3.3}$ ($3.3)\%$}\tabularnewline
\hline
\hline 
\multicolumn{1}{|c|}{}&
$A$ {[}pb] &
1.96&
2.01&
2.058&
248&
248.4&
249.1\tabularnewline
\multicolumn{1}{|c|}{$T=t$}&
$B$ {[}pb$\cdot$GeV$^{-1}$] &
-0.034&
-0.036&
-0.037&
-1.93&
-2.19&
-2.24\tabularnewline
\cline{2-2} \cline{3-3} \cline{4-4} \cline{5-5} \cline{6-6} \cline{7-7} \cline{8-8} 
\multicolumn{1}{|c|}{($t$-channel)}&
$\Delta_{\mu}(m_{t}=171)$ &
-2.7\%&
reference &
2.6\%&
-1.6\%&
reference &
2.4\%\tabularnewline
\cline{2-2} \cline{3-5} \cline{6-8} 
\multicolumn{1}{|c|}{}&
$\Delta_{PDF}(m_{t}=171)$ &
\multicolumn{3}{c|}{\textbf{10.3\%}}&
\multicolumn{3}{|c|}{3.2\%}\tabularnewline
\hline
\hline 
&
$A$ {[}pb] &
1.013&
0.967&
0.925&
11.83&
11.710&
11.67\tabularnewline
\multicolumn{1}{|c|}{$T=t$}&
$B$ {[}pb$\cdot$GeV$^{-1}$] &
-0.025&
-0.024&
-0.023&
-0.248&
-0.247&
-0.248\tabularnewline
\cline{2-2} \cline{3-3} \cline{4-4} \cline{5-5} \cline{6-6} \cline{7-7} \cline{8-8} 
\multicolumn{1}{|c|}{($s$-channel)}&
$\Delta_{\mu}(m_{t}=171)$&
+5\%&
reference&
-4\%&
+1.0\%&
reference &
-0.4\%\tabularnewline
\cline{2-2} \cline{3-5} \cline{6-8} 
\multicolumn{1}{|c|}{}&
$\Delta_{PDF}(m_{t}=171)$ &
\multicolumn{3}{c|}{3.4\%}&
\multicolumn{3}{|c|}{3.0\%}\tabularnewline
\hline
\end{tabular}
\end{table}

\begin{table}[H]

\caption{Correlation cosines $\cos\varphi$ between single-top, $W,$ $Z,$
and $t\bar{t}$ cross sections at the Tevatron Run-2 (Tev2) and LHC,
computed with CTEQ6.6 PDFs. \label{tab:CorrsSingleTop}}

\begin{tabular}{|c||c|c|c||c|c|c|}
\hline 
Single-top&
\multicolumn{6}{c|}{Correlation $\cos\varphi$ with}\tabularnewline
production channel&
$Z^{0}$ (Tev2)&
$W^{\pm}$(Tev2)&
$t\bar{t}$ (Tev2)&
$Z^{0}$ (LHC)&
$W^{\pm}$ (LHC)&
$t\bar{t}$ (LHC)\tabularnewline
\hline
\hline 
$t-$channel (Tev2)&
-0.18&
-0.22&
\textbf{0.81}&
\textbf{-0.82}&
\textbf{-0.79}&
0.56\tabularnewline
\hline 
$t-$channel (LHC)&
0.09&
0.14&
-0.64&
0.56&
0.53&
-0.42\tabularnewline
\hline 
$s-$channel (Tev2)&
\textbf{0.83}&
\textbf{0.79}&
0.18&
0.22&
0.27&
-0.3\tabularnewline
\hline 
$s-$channel (LHC)&
\textbf{0.81}&
\textbf{0.85}&
-0.42&
0.6&
0.68&
-0.33\tabularnewline
\hline
\end{tabular}
\end{table}

\begin{table}[H]

\caption{CTEQ6.6M total cross sections $\sigma$ and PDF errors $\Delta\sigma$
for Higgs boson production at $\sqrt{s}=14\mbox{ TeV}$ shown in Fig.~\ref{fig:Correlation-cosine-vs-mass}.
\label{tab:HiggsXSecsHC}}

\begin{tabular}{|c|c|c|c|}
\hline 

Scattering&
\multicolumn{3}{c|}{$\sigma$ and $\Delta\sigma$ (pb)}\tabularnewline
process&
$m{}_{h}=120$ GeV&
$m_{h}=200$ GeV&
$m_{h}=500$ GeV\tabularnewline
\hline
\hline 
$pp\rightarrow(gg\rightarrow h^{0})X$&
33(1)&
14.0(4)&
4.0(2)\tabularnewline
$pp\rightarrow(b\bar{b}\rightarrow h^{0})X$&
2750(130)&
460(20)&
11.0(7)\tabularnewline
$pp\rightarrow(c\bar{s}+c\bar{b}\rightarrow h^{+})X$&
16(1)&
2.61(14)&
0.063(3)\tabularnewline
$pp\rightarrow W^{+}h^{0}X$&
1.15(3)&
0.201(6)&
0.0062(3)\tabularnewline
$pp\rightarrow W^{-}h^{0}X$&
0.74(2)&
0.117(4)&
0.00292(16)\tabularnewline
$pp\rightarrow(WW\rightarrow h^{0})X$&
2.80(8)&
1.60(5)&
0.36(1)\tabularnewline
\hline
\end{tabular}
\end{table}
\begin{table}[H]

\caption{Relative differences $\Delta_{GM}\equiv\sigma_{6.1}/\sigma_{6.6}-1$
between CTEQ 6.1 and CTEQ 6.6 cross sections for Higgs boson production
at the LHC, compared to the PDF uncertainties $\Delta_{PDF}$ in these
processes. The $Ah^{\pm}$ cross section is for combined production
of positively and negatively charged Higgs bosons, with $m_{h}$ being
the mass of the CP-odd boson ($m_{h}=m_{A}),$ and $m_{h^{\pm}}$
given by $m_{h\pm}^{2}=m_{A}^{2}+M_{W}^{2}$\textbf{.} \label{tab:Higgsproduction}}

\begin{tabular}{|c||c|c||c|c||c|c||c|c||c|c||c|c||c|c|}
\hline 
$m_{h}$&
\multicolumn{14}{c|}{$\Delta_{GM}(\%)|\Delta_{PDF}(\%)$}\tabularnewline
\cline{2-3} \cline{4-5} \cline{6-7} \cline{8-9} \cline{10-11} \cline{12-13} \cline{14-15} 
\multicolumn{1}{|c||}{(GeV)}&
\multicolumn{2}{c||}{VBF}&
\multicolumn{2}{c||}{$Z^{0}h$}&
\multicolumn{2}{c||}{$Ah^{\pm}$}&
\multicolumn{2}{c||}{$gg\rightarrow h$}&
\multicolumn{2}{c||}{$c\bar{b}\rightarrow h^{+}$}&
\multicolumn{2}{c||}{$c\bar{s}\rightarrow h^{+}$}&
\multicolumn{2}{c|}{$c\bar{s}+c\bar{b}\rightarrow h^{+}$}\tabularnewline
\hline 
100&
\textbf{-3.8}&
\textbf{3.1}&
\textbf{-3.2}&
\textbf{2.7}&
\textbf{-3.2}&
\textbf{4.3}&
0.6&
4.4&
1.5&
5.9&
\textbf{-18}&
\textbf{10}&
\textbf{-8.4}&
\textbf{6.9}\tabularnewline
\hline 
200&
-1.8&
2.8&
-1.6&
2.8&
-1.9&
 4.3&
1.7&
3.2&
 2.1&
4.7&
\textbf{-16}&
\textbf{8}&
\textbf{-6.6}&
\textbf{5.4}\tabularnewline
\hline 
300&
-1.6&
2.8&
-0.6&
3&
-0.4&
5.3&
2.3&
2.7&
1.9&
 4.3&
 \textbf{-14}&
\textbf{7}&
\textbf{-6.2}&
\textbf{4.5}\tabularnewline
\hline 
400&
-0.1&
3.3&
0&
3.4&
0.7&
 6.6&
2.8&
3.8&
2&
4.8&
\textbf{-13}&
\textbf{6.3}&
\textbf{-5.6}&
\textbf{4.4}\tabularnewline
\hline 
500&
0.2&
2.8&
0.4&
3.7&
1.1&
 7.6&
3.3&
3.9&
2.3&
6.1&
\textbf{-12}&
\textbf{6.3}&
\textbf{-5}&
\textbf{5.1}\tabularnewline
\hline 
600&
-0.7&
3.5&
0.7&
4.1&
1.6&
9.2&
3.8&
5.0&
2.8&
 8&
 \textbf{-11}&
 \textbf{6.8}&
-4.2&
6.4\tabularnewline
\hline 
700&
0.2&
3.0&
0.9&
4.4&
2.1&
11&
4.3&
6.3&
3.4&
10&
\textbf{-9.9}&
\textbf{7.7}&
 -3.4&
8\tabularnewline
\hline 
800&
2.3&
3.5&
1&
4.8&
2.8&
13&
4.9&
7.8&
4.1&
12&
\textbf{-8.7}&
\textbf{9}&
-2.4&
10\tabularnewline
\hline
\end{tabular}
\end{table}

\newpage

\end{document}